\documentclass[twocolumn,aps,pra,showpacs,floatfix,floats]{revtex4}
\usepackage{times}
\usepackage{subfigure}
\usepackage{graphicx}
\usepackage{amssymb}
\usepackage{bm}

\makeatletter

\begin{document}

\title{Hamiltonian Formulation of Quantum Error Correction
                  and Correlated Noise:\\ 
The Effects Of Syndrome Extraction in the Long Time Limit}

\author{E. Novais,$^{1}$ Eduardo R. Mucciolo,$^{2}$ and Harold
U. Baranger$^{1}$}

\affiliation{$^{1}$Department of Physics, Duke University, Box 90305, 
Durham, North Carolina 27708-0305, USA\\
$^{2}$Department of Physics, University of Central
Florida, Box 162385, Orlando, Florida 32816-2385, USA}

\date{\today}

\begin{abstract}
We analyze the long time behavior of a quantum computer running a
quantum error correction (QEC) code in the presence of a correlated
environment. Starting from a Hamiltonian formulation of realistic
noise models, and assuming that QEC is indeed possible, we find formal
expressions for the probability of a given syndrome history and the
associated residual decoherence encoded in the reduced density
matrix. Systems with non-zero gate times (``long gates'') are included
in our analysis by using an upper bound on the noise. In order to
introduce the local error probability for a qubit, we assume that
propagation of signals through the environment is slower than the QEC
period (hypercube assumption). This allows an explicit calculation in
the case of a generalized spin-boson model and a quantum frustration
model. The key result is a dimensional criterion: If the correlations
decay sufficiently fast, the system evolves toward a stochastic error
model for which the threshold theorem of fault-tolerant quantum
computation has been proven. On the other hand, if the correlations
decay slowly, the traditional proof of this threshold theorem does not
hold. This dimensional criterion bears many similarities to criteria
that occur in the theory of quantum phase transitions.
\end{abstract}


\maketitle

\section{Introduction}
\label{sec:intro}

Quantum computation provides a fundamentally new way to process data;
as a theory, it is complete and remarkably rich \cite{NC00}. However,
any real quantum computer is subject to an implacable physical
reality: components of a computer will always be faulty due to
environmental noise. Hence, the builder of a quantum computer faces
the conundrum of having to isolate the device from its surroundings
and, simultaneously, of needing to act on it and read its output
\cite{UNR95}. Many strategies have been devised to address this
problem \cite{NC00,LCW98,VL98,VKL00,LS05,SV06}, the most general being
quantum error correction \cite{NC00,Ste96c,Ste96,CS96,CRS+98}.

Quantum error correction (QEC) should be understood as a perturbative
approach \cite{KL97}, where one can estimate the probability of having
an ``error'' in the wave function of the quantum computer after a
certain time. It is naturally formulated as a perturbation expansion
in powers of the coupling between the computer and the environment
\cite{KL97}.
QEC cannot, in general, perfectly correct the quantum evolution, and
the interference of the amplitudes for the various processes that
occur implies that quantum information is always lost to the
environment \cite{KL97}. However, as we discuss below, QEC can very
effectively slow down this loss. In fact, a central theoretical result
is the ``threshold theorem'': it states that if the error probability
is smaller than a critical value, quantum computation can be sustained
indefinitely
\cite{Got98,Pre,KLZ98b,DAMB99,Got99,Aha00,KLZ01,Ste03,Kni05b}. The
word ``indefinitely'' deserves some clarification: For the problems
that we discuss, it means that given a calculation and a desired
precision, it is always possible to construct a quantum circuit that
will provide the correct result with high enough probability.

QEC has been largely developed using phenomenological ``error
models''. Rarely is a connection to a microscopic quantum dynamical
system found in the literature (see, however,
Refs.~\cite{AHH+02,CSG04b,KF05,ALZ06}). In contrast, here we pursue
exactly such a connection: We discuss the formal steps needed to link
the theory of error correction with microscopic Hamiltonian
models. Furthermore, because of the perturbative nature of the method,
it is possible to draw a close parallel between the ``threshold
theorem'' and the theory of quantum phase transitions. We find that if
a certain inequality holds, an error threshold always exists. When the
inequality is not satisfied, either a new version of the threshold
criterion is required or fault tolerant quantum computation is not
possible at all. For the moment, we are not able to distinguish
between these two possibilities.

Our analysis is based on the following assumptions. First and
foremost, we assume that it is possible to perform the building blocks
of quantum error correction, namely, preparation of states, quantum
gates, and measurements. Second, we consider that the environment is
described by a free field theory in which thermal fluctuations can be
effectively suppressed. Finally, the main simplifying assumption of
our discussion is that the qubits are sufficiently separated in space
for an entire error correction procedure to be performed before
correlations between nearby qubits develop.  The probability of an
error in an individual qubit within a QEC cycle is, therefore,
independent of all other qubits. This does not imply that there are no
spatial correlations; rather, they develop on longer time scales,
while the error correction procedure is done faster than a certain
characteristic time. We emphasize that this hypothesis is not a
limitation of the general theoretical framework that we describe, but
simply a way to connect to the traditional proofs of the ``threshold
theorem'' in terms of stochastic error models.

The paper is organized as follows. Because of the interdisciplinary
nature of the subject, this Introduction continues with a discussion
of two points. First, the difficulties in taking into account
correlations in the environment are explained in
Sec.~\ref{sec:the-problem-of} from a perturbative point of view. Then,
in Sec.~\ref{sec:QEC}, we discuss the QEC method from a physics
viewpoint and present some results for the standard stochastic error
model.  We start the body of the paper by developing the relation
between error models and quantum codes
(Sec.~\ref{sec:perturbation}). The key issue of QEC in a correlated
environment is treated in Sec.~\ref{sub:QEC-and-perturbation}. Our
main results delineating when the perturbative treatment is valid
appear in Sec.~\ref{sec:hypercube}. At the end of this Section, we
provide a brief comparison between our results and those of
Ref.~\onlinecite{AKP06}. Sec.~\ref{sec:QPT} discusses parallels
between the threshold theorem of QEC and the theory of quantum phase
transitions. Finally, in Sec.~\ref{sec:conclusions} we summarize our
results and comment on some open problems.

\subsection{The problem of correlated environments}
\label{sec:the-problem-of}

In order to set the stage for the analysis in the presence of QEC, we
first look at the problem of errors created by a correlated
environment in an unprotected system. In the Schr{\"o}dinger equation
governing the time evolution of a quantum system,
the Hamiltonian $H$ can usually be separated into a single-particle
term $H_{0}$ and a many-particle interaction part $V$. A formal
solution of this equation is given by the Dyson series in the
interaction picture. Solution by iteration shows that the time
evolution operator is
\begin{equation}
U\left( t, 0 \right) = T_{t}\, e^{-\frac{i}{\hbar}
\int_{0}^{t}dt^{\prime}\, V \left( t^\prime \right)},
\label{eq:evol-int-1}
\end{equation}
with $T_{t}$ denoting the time ordering operator and $V(t) =
e^{\frac{i}{\hbar} H_{0}t} V e^{-\frac{i}{\hbar}H_{0}t}$.  If $V$
represents the interaction between the quantum computer and its
surroundings, each insertion of $V$ in Eq.~(\ref{eq:evol-int-1})
corresponds to an ``error'' in the computer evolution. Hence,
Eq.~(\ref{eq:evol-int-1}) provides the natural framework to study the
effects of the environment on the state of the quantum computer.

It is always possible to give an upper bound to the ``error
probability'' \cite{KLV00}. The reason is that Dyson's series is
absolutely convergent for finite times and bounded operators (see
Appendix \ref{sec:absolute-convergence}). In short, the bounding is
done by defining the ``sup'' operator norm and the evolution operator
with at least one ``error'' (one insertion of $V$),
\begin{eqnarray}
\mathcal{E}\left(t\right) & = & U(t,0)-1 \nonumber \\ & = &
-\frac{i}{\hbar}\int_{0}^{t} dt^{\prime}\, V \left( t^{\prime}
\right)\, U( t^{\prime}, 0) \;.
\end{eqnarray}
The norm of $\mathcal{E}$ is related to the probability of having
errors in the computer. The calculation is simple and yields
\begin{equation}
\label{eq:Fbound}
\left|\left|\mathcal{E}\left(t\right)\right|\right|  \leq 
\frac{1}{\hbar}\int_{0}^{t}dt^{\prime}\, \left| \left| V \left(
t^{\prime} \right) \right|\right| \leq  \frac{\Lambda
t}{\hbar} \;,
\end{equation} 
where we used the triangular inequality, the unitarity of $U$, and
defined $\Lambda$ as the largest eigenvalue of $V$ (with corresponding
eigenvector $\Psi_{\Lambda}$). One can understand this bound as simply
a restatement of $\left|\sin x\right| \leq \left|x\right|$, as
follows:
\begin{eqnarray}
\label{eq:FdagF}
\mathcal{E}^{\dagger}(t)\, \mathcal{E}(t) & = & \left[ 2-U^{\dagger}
 (t) - U (t) \right] \nonumber \\ & = & 2 \left[ 1-T_{t}
 \cos\frac{1}{\hbar} \int_{0}^{t} dt^{\prime} V (t^{\prime}) \right]
\end{eqnarray}
so
\begin{eqnarray}
\sqrt{\left\langle \Psi_{\Lambda} \right| \mathcal{E}^{\dagger}(t) \,
\mathcal{E}(t) \left| \Psi_{\Lambda} \right\rangle} & = & \sqrt{2
\left( 1- \cos \frac{\Lambda t}{\hbar}\right)} \nonumber \\ & = &
\left| \sin \frac{\Lambda t} {2\hbar} \right| \leq \left|
\frac{\Lambda t}{2\hbar} \right|.
\end{eqnarray}

The norm $\left|\left|\mathcal{E}\right|\right|$ has been very useful
in problems involving non-Markovian noise
\cite{KLV00,TB05,AGP06,Rei06,AKP06}. However, in QEC, an analysis
based on the bound Eq.~(\ref{eq:Fbound}) only makes sense when $\left|
\left| \mathcal{E} \right| \right| \ll 1$, while we are concerned with
the long time limit, $|\Lambda t| \gg 1$, for which this bound on the
norm of the error diverges. In this case, Dyson's series is only
asymptotically convergent and the ``sup'' norm is of no practical
use. Hence, it is important to express the error probability
differently.

We must go back full circle and reexamine the Dyson series for the
time evolution of a particular state, instead of the worst case
scenario explored by the ``sup'' norm approach. Henceforth, we will be
mainly interested in an interaction Hamiltonian with the general form
\begin{equation} 
V (t) = \lambda \int_{0}^{L}d{\bf x}\, f \left({\bf x},t\right),
\end{equation}
where $\lambda \ll 1$ is a coupling constant, $L$ is the size of the
system, and $f$ is some function of the degrees of freedom of a free
theory whose Hamiltonian is $H_{0}$. Because we are interested in
correlated non-Markovian noise, we assume that the free fields are
such that the asymptotic expression for the
two-point correlation function is a power law,
\begin{equation}
\left\langle \Psi \left| f\left({\bf x}_{1},t_{1}\right) f\left({\bf
x}_{2},t_{2}\right) \right| \Psi \right\rangle \sim \mathcal{F}
\left(\frac{1}{\left(\Delta x\right)^{2\delta}},\frac{1} {\left(
\Delta t\right)^{2\delta/z}} \right), \label{general2pointfunction}
\end{equation}
where $\Delta x = |{\bf x}_{1} - {\bf x}_{2}|$ and $\Delta t =
|t_1-t_2|$\cite{asymptoticform}.  Here, $\delta$ is the scaling
dimension of $f$, $z$ is the so-called dynamical exponent, and $\left|
\Psi \right\rangle$ is a fixed eigenstate of $H_{0}$ (which we will
usually take to be the ground state of the environment).

The motivation for developing a perturbative expansion of the
evolution operator (the Dyson series in the interaction picture) is
the hope that a few terms in the series or a summable family of them
will capture most of the physics. It is then assumed that small
coupling can guarantee fast convergence. However, since $\left| \left|
\mathcal{E} \right|\right|$ is not necessarily small, the number of
terms that contribute substantially to the series can grow faster that
the smallness of consecutive terms. In order to see that, let us
calculate the probability of an evolution with errors using
Eq.~(\ref{eq:FdagF}),
\begin{equation}
\frac{\left\langle \Psi\right |\mathcal{E}^{\dagger} \left(t\right) 
\mathcal{E}
\left(t\right) \left| \Psi\right\rangle }{2} = 1 - \left\langle \Psi
\right| T_{t} \cos \left[ \frac{1}{\hbar} \int_{0}^{t} dt^{\prime}
V(t^{\prime}) \right] \left| \Psi \right\rangle.
\end{equation}
Since we are assuming a non-interacting free Hamiltonian, we can use
Wick's theorem. It is then straightforward to show that there is at
least one term at each order $m$ in the series that contributes
``extensively'' as $\sim\! \lambda^{2m} (Lt)^{2m (D+z-\delta)}$.  A
simple example is given by the series of ``bubble'' diagrams, where
the $m^{th}$ order term is given by the contractions
\begin{equation}
\int_{0}^{t} dt_1 ... \int_{0}^{t_{m-1}} dt_m \left\langle V(t_1)
V(t_2) \right\rangle ... \left\langle V(t_{m-1}) V(t_m) \right\rangle.
\end{equation}
Disregarding numerical prefactors unimportant for our discussion, we
sum the series as a geometric progression to obtain
\begin{equation}
\frac{\left\langle \Psi\right|F^{\dagger}(t) F(t) \left| \Psi
\right\rangle}{2} \sim \frac{\lambda^{2}\, (Lt)^{2(D+z-\delta)}} {1 +
\lambda^{2}\, (Lt)^{2(D+z-\delta)}}.
\end{equation}
Therefore, for $D+z-\delta>0$ there is no guarantee that the
perturbation series converges. Conversely, if $D+z-\delta<0$,
higher-order terms in the series should be increasingly less
important.  Thus, for $D+z-\delta>0$ the probability of an evolution
with ``errors'' tends to one, whereas for $D+z-\delta<0$ it will
depend only on the ``non-extensive'' terms in the series. The same
analysis can be immediately transported to the study of the fidelity
$\left| \left\langle \Psi \right| U(t) \left| \Psi \right\rangle
\right|$, where we see that for a relevant perturbation,
$D+z-\delta>0$, the overlap between the initial state and the evolving
wave function tends to zero (an orthogonality catastrophe). This sort
of ``infrared'' problem provides a contact point with the theory of
quantum phase transitions, where the same kind of considerations also
appear when calculating the partition function using the imaginary
time formalism (see Appendix \ref{sec:Perturbative-expansion-in}).

In the body of this paper, our main goal is to transfer these ideas of
relevance and irrelevance of a perturbation to the evolution of a
quantum computer protected by QEC.

\subsection{Quantum error correction}
\label{sec:QEC}

Quantum error correction is arguably the most versatile method to
protect quantum information from decoherence \cite{Zur03}. It is a
clever use of two features of quantum mechanics: entanglement and (in
its traditional form) wave packet reduction due to measurement. Thus,
before we start our discussion of QEC, it is important to carefully
define what we mean by entanglement and decoherence.

An entangled state of two quantum systems is a state that cannot be
described as a direct tensor product of states of individual systems
or probabilistic mixtures of tensor-product states. As an example,
consider two physical qubits (hereafter referred to by the subscripts
$1$ and $2$). Each qubit has a Hilbert space isomorphic to a complex
projective plane of dimension one, $\mathbb{CP}^{1}$ (see Appendix
\ref{sec:hilbert} for details). However, the combined Hilbert space is
not isomorphic to $\mathbb{CP}_{(1)}^{1} \times
\mathbb{CP}_{(2)}^{1}$, but to the much larger $\mathbb{CP}^{3}$. All
states in $\mathbb{CP}^{3}$ outside $\mathbb{CP}_{\left(1\right)}^{1}
\times \mathbb{CP}_{\left(2\right)}^{1}$ are said to be entangled. An
important subtlety is the implicit notion of a preferred
``basis''. Although we can choose from an infinite number of
$\mathbb{CP}^{1} \times \mathbb{CP}^{1}$ subspaces inside the same
$\mathbb{CP}^{3}$, nature gives us a natural choice, namely,
$\mathbb{CP}_{\left(1\right)}^{1} \times
\mathbb{CP}_{\left(2\right)}^{1}$.

In the working of a quantum computer, entanglement has two opposite
roles. On the one hand, entanglement between qubits is the key element
in a quantum computation that distinguishes it from its classical
counterpart \cite{LP01}. On the other hand, when the computer and the
environment become entangled, precious quantum information is
lost. Usually, the latter effect is referred to as decoherence. In the
literature, there are two different definitions of decoherence. In a
strict sense, decoherence is the decay in time of the coherences
(off-diagonal elements of the reduced density matrix), while
dissipation involves the exchange of energy with the environment and
changes in populations (the diagonal terms of the density
matrix). However, the word ``decoherence'' is also used in a broader
sense involving changes in both diagonal and off-diagonal entries of
the density matrix. In this paper we choose the latter use of the
word. The reason is that from a quantum error correction perspective
changes in diagonal and off-diagonal entries are ``dual'' to each
other \cite{NC00}.

There is a simple heuristic explanation for error correction: Usually,
noise is regarded as a local phenomenon, thus its damaging effect in
the computer should be less pronounced if the information is
delocalized among several qubits. This is precisely how classical
error correction codes work. A simple example of the latter is a
majority vote, where the information of a bit is copied into three
physical bits, $0 \!\rightarrow\! 000$ and $1 \!\rightarrow\! 111$.
If the probability of an error in a given qubit is $\epsilon$, the
probability of having two independent errors, and consequently a total
information loss, is $\epsilon^{2} \ll \epsilon$. Thus encoding
increases the level of protection of the information.

It is tempting to start explaining QEC from this perspective. However,
the no-cloning theorem \cite{NC00} states that it is impossible to
copy an unknown quantum state. The alternative approach is to use an
entangled state involving two or more qubits to store the quantum
information. This clearly delocalizes the information, but it is at
odds with the intuitive notion that entangled states are in general
more fragile to the effects of the environment (this intuition is
driven by the quantum-to-classical transition due to decoherence, see
Appendix \ref{sec:SB-decoherence} for a concrete example). Thus,
delocalizing the information using entanglement does not alone solve
the problem. It is possible to use unitary operations to transfer the
entanglement between the qubits and the environment to a constant
fresh supply of ancilla qubits \cite{NC00,Preskill-notes}.  However,
it is more traditional in QEC to use the partial measurements of
\emph{some} ancilla qubits to reduce the quantum interference with the
environment \cite{NC00}. Measurements here have to be understood as
the projection of the state of one of the qubits (an ancilla) onto a
certain basis or reference state. The outcome of this projection is a
classical bit (``zero'' or ``one'') and is called a syndrome.  The
partial wave packet reductions caused by syndrome extraction steer the
long-time evolution of the quantum computer.  Recently, it has been
shown that the duration of the measurement is not fundamental to the
QEC procedure \cite{DA07}. In fact, this process can be quite long
without jeopardizing the method.

\begin{figure}[t]
\includegraphics[width=7cm]{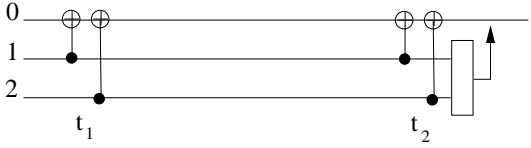}
\caption{A 3 qubit quantum error correction (QEC) code
\cite{NC00,Ste96,Ste96b,CS96,CRS+97}. The initial wave function,
$\left|\psi_{0}\right\rangle\otimes ( \left|\uparrow\right\rangle
\!+\! \left|\downarrow\right\rangle )/2 \otimes (
\left|\uparrow\right\rangle \!+\! \left|\downarrow\right\rangle )/2$,
is encoded by two controlled-NOT (CNOT) gates, $R_{\rm CNOT} =
\sigma_{i}^{-}\sigma_{i}^{+}\sigma_{j}^{x} \!+
\sigma_{i}^{+}\sigma_{i}^{-}$, into an entangled state
$\left|\psi_{\rm encode}\right\rangle
=\alpha\left|\bar{\uparrow}\right\rangle +
\beta\left|\bar{\downarrow}\right\rangle $ with
$\left|\bar{\uparrow}\right\rangle
=\left(\left|\uparrow\uparrow\uparrow\right\rangle
+\left|\uparrow\downarrow\downarrow\right\rangle
+\left|\downarrow\uparrow\downarrow\right\rangle
+\left|\downarrow\downarrow\uparrow\right\rangle \right)/2$ and
$\left|\bar{\downarrow}\right\rangle
=\left(\left|\downarrow\downarrow\downarrow\right\rangle
+\left|\downarrow\uparrow\uparrow\right\rangle
+\left|\uparrow\downarrow\uparrow\right\rangle
+\left|\uparrow\uparrow\downarrow\right\rangle \right)/2$. After some
time, it is decoded by a second pair of CNOT gates. An error in
$\left|\psi\right\rangle $ is identified by measuring the values of
$\sigma_{2}^{x}$ and $\sigma_{3}^{x}$ (rectangle). The QEC cycle ends
with the correction of a possible phase-flip (arrow).}
\label{fig:3-qubit-code}
\end{figure}

\begin{figure}[t]
\includegraphics[width=6cm]{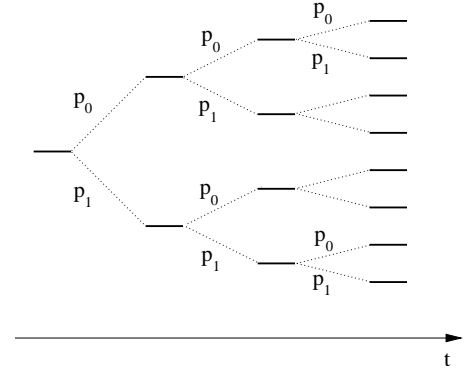}
\caption{A syndrome history diagram. Each solid line represents the
evolution of a logical qubit. At the end of a QEC cycle, a phase flip
error is detected or not with probabilities $p_{1}$ and $p_{0}$,
respectively.  A path provides the history of the logical qubit and is
recorded as a sequence of syndromes.}
\label{fig:Fault-path}
\end{figure}

A simple example illustrates how QEC works
\cite{Ste96,CS96,NC00}. Suppose that we have an error model consisting
of independent baths for each qubit which can cause only phase errors,
and an initial qubit in the state $\left|\psi_{0}\right\rangle
=\alpha\left|\uparrow\right\rangle +\beta\left|\downarrow\right\rangle
$ that we want to protect. The 3-qubit code provides the simplest
error correction procedure for this problem.  In
Fig.~\ref{fig:3-qubit-code}, we define the encoding/decoding methods
in a QEC cycle.  At the end of a cycle, the probability of measuring
the syndrome of a phase flip error in one of the three physical qubits
is \cite{NB06}
\begin{equation}
p_{1} = 3\epsilon,
\end{equation}
and the probability of the syndrome indicating no error in the logical
qubit is
\begin{equation} 
p_{0} = 1-p_{1}.
\end{equation}
The residual decoherence that can \emph{not} be corrected by the QEC
procedure is closely related to these probabilities. In the case of a
cycle in which the syndrome indicates that one error occurred in any
of the physical qubits, dephasing of the logical qubit is given by the
reduction of the off-diagonal density matrix element \cite{NB06},
\begin{equation}
\rho_{\bar{\uparrow}\bar{\downarrow}}^{(1)} \approx \alpha \beta^\ast
\left(1-2\epsilon\right),
\end{equation}
while for a cycle with a syndrome indicating no error, the dephasing
is weaker,
\begin{equation}
\rho_{\bar{\uparrow}\bar{\downarrow}}^{(0)} \approx \alpha \beta^\ast
\left(1-2\epsilon^{3}\right).
\end{equation}
After $N$ of these cycles, the probability of having $m$ uncorrelated
errors is
\begin{equation}
\label{eq:Pm}
\mathcal{P}_{m} = \left( \begin{array}{c} N \\ m \end{array} \right)
p_{0}^{N-m}p_{1}^{m},
\end{equation}
with an associated residual decoherence of
\begin{equation}
\label{eq:rhoud}
\rho_{\bar{\uparrow} \bar{\downarrow}}^{(m)} \approx \alpha \beta^\ast
\left(1 - 2\epsilon^{3}\right)^{N-m}
\left(1-2 \epsilon\right)^{m}.
\end{equation}
An elegant visualization of these events is given by a ``syndrome
history diagram'' of Fig.~\ref{fig:Fault-path} (see for instance
Ref.~\onlinecite{TB05} for a similar discussion). An ordered set of
syndromes labels a particular evolution of the logical qubit. From the
syndrome history one can find the most likely evolution and the
associated residual decoherence. For our 3-qubit code example, the
most likely evolution is given by the mean value of $m$,
$\bar{m}=Np_{1}$. Thus, the residual decoherence of the logical qubit
is given by
\begin{equation} 
\rho_{\bar{\uparrow}\bar{\downarrow}} \approx \alpha \beta^\ast\,
e^{-6 N\epsilon^{2}}.
\end{equation}
Therefore, as long the number of QEC cycles $N\ll\epsilon^{-2}$, the
probability of measuring the correct initial state of the logical
qubit is very high. We can quantify the amount of information that is
lost by calculating the von Neumann entropy
$S=-\mbox{tr}\left(\rho\ln\rho\right)$:
\begin{eqnarray}
 \lim_{N\ll\epsilon^{-2}}S & \approx & 12N |\alpha|^{2} |\beta|^{2}
 \epsilon^{2} \left[ 1 - \ln \big( 12N |\alpha|^{2} |\beta|^{2}
 \epsilon^{2} \big) \right]  \quad\;\\ 
\lim_{N\gg\epsilon^{-2}}S & \approx
 & - |\alpha|^{2} \ln |\alpha|^{2}- |\beta|^{2} \ln |\beta|^{2} \;.
\end{eqnarray}
Note that the loss of information can be substantial if the number of
cycles is so large that $N\gg\epsilon^{-2}$.

If the information needs to be protected for a long period of time, we
have to modify the protection scheme. The most straightforward
approach is to consider a concatenated circuit where each qubit in
Fig.~\ref{fig:3-qubit-code} is a logical qubit itself and each gate is
a logical gate, resulting in an effective reduction of $p_{1}$. Layers
and layers of protection can be added as needed \cite{NC00,AGP06}. A
chief concern when applying this approach is whether the steps
required in the addition of more qubits and operations do not actually
increase the chance of errors (since they increase the combinatorial
factors in the probability distribution). This question is addressed
by fault-tolerant quantum computation theory
\cite{Got98,KLZ98b,Got99,DAMB99,KLZ01,AGP06}, which has as its main
result the so-called threshold theorem: If the ``noise strength''
$\epsilon$ is smaller than a certain critical value, then the
introduction of an additional layer of concatenation improves the
protection of the information.

A key ingredient in the derivation of the noise threshold is the
assumption that a probabilistic structure similar to the one that we
outlined above exists. Here rests the main concern of this
paper. There are many physical situations where an environment can
induce strong memory effects and spatial correlations among
qubits. Hence, it may not be obvious how to define the ``error
probabilities'' of a qubit. This hinders the traditional theory of QEC
and threshold analysis, thus motivating a careful study of the
dynamics of quantum computers protected by QEC.

\section{Error Models and Quantum Codes}
\label{sec:perturbation}

The syndrome history used to describe the logical qubit history can be
converted into a more formal description of the computer dynamics.  In
our discussion, we will assume an environment, $H_{0}$, described by a
free field theory with an ultraviolet cutoff $\Lambda$, a
characteristic wave velocity $\mathtt{v}$, and a dynamical exponent
$z$. Although simple, a free field theory faithfully represents many
physically relevant environments: the electromagnetic field, a phonon
field, spin waves, a bosonic bath, or, more generally, any two-body
direct interactions between qubits that was split by a
Hubbard-Stratanovich field. In addition, we include in the Hamiltonian
a term to account for the sequence of quantum gates performed on the
qubits, $H_{\rm QC}(t)$. Hence, the total Hamiltonian is
\begin{eqnarray}
H\left(t\right) & = & H_{0} + H_{\rm QC}(t) + V.
\label{eq:totalHamil}
\end{eqnarray} 
The interaction term will be assumed to have the form of a vector
coupling between qubits and the environment,
\begin{equation}
V = \sum_{{\mathbf{x}}}\sum_{\alpha=\{ x,y,z\}}
\frac{\lambda_{\alpha}}{2} f_{\alpha}({\mathbf{x}})\,
\sigma^{\alpha}({\mathbf{x}}),
\end{equation}
where $\vec{\sigma}({\mathbf{x}})$ are Pauli matrices for the qubit
located at ${\mathbf{x}}$, $\lambda_{\alpha}$ are the coupling
strengths, and $f_{\alpha}({\mathbf{x}})$ are functions of the
environment operators \cite{static-imperfections}.  Since
$\left[H_{0},H_{\rm QC}\right]=0$, we adopt an interaction picture
that follows not only the environment but also the evolution of the
computer (see Appendix \ref{sec:Interaction-picture}). In this
rotating frame, the evolution operator is
\begin{equation}
U (t,0) = T_{t}\ e^{-\frac{i}{\hbar} \int_{0}^{t} dt^{\prime} V
\left(t^{\prime}\right)} \;.
\label{eq:unitaryevolution}
\end{equation}

The interaction $V(t)$ depends on the quantum code and its
implementation. Nevertheless, there are two possible ways to keep
the discussion code independent: 

(i) In our previous work \cite{NB06,NMB07}, we assumed that quantum
gates were performed faster than the environment response time (which
is of order the inverse of the ultraviolet cutoff frequency
$\Lambda$). We call this approximation the ``fast gate'' limit. For
this case, we have the evolution of the computer between gates given
by
\begin{equation}
V(t) = \sum_{{\mathbf{x}}} \sum_{\alpha=\{ x,y,z\}}
\frac{\lambda_{\alpha}}{2} f_{\alpha}({\mathbf{x}},t) \,
\sigma^{\alpha}({\mathbf{x}}),
\label{eq:fastgates}
\end{equation}
with $f_{\alpha}({\mathbf{x}},t) = e^{\frac{i}{\hbar} H_{0}t}
f_{\alpha} ({\mathbf{x}}) e^{-\frac{i}{\hbar}H_{0}t}$. Then, when a
gate is performed the action on the qubit is instantaneous and the
subsequent evolution is once again governed by
Eq.~(\ref{eq:fastgates}). 

(ii) A second possibility is to derive an upper bound on the effects
of correlations. In order to do that, we must first discuss how slow
gates, which are performed over time intervals larger than
$\tau_{c}=1/\Lambda$, change Eq.~(\ref{eq:fastgates}). Then, we can
define an effective interaction $V_{\rm eff}$ that takes into account
the slowness of the gates and serves as an upper bound to the exact
(and code-dependent) $V$. Clearly, the real experimental situation
rests between the two limits (i) and (ii).

Before we begin a detailed description of how to handle case (ii), let
us note that here the terminology ``fast'' and ``slow'' gates follows
the QEC literature: Fast (slow) gates have a duration much shorter
(longer) than $\tau_c$. However, as will become clear later, the
relevant time scale that appears in the study of correlation effects
is the period or duration of the error correction cycle,
$\Delta$. Thus, in that context, short (``fast'') or long (``slow'')
dynamical effects will be naturally defined with respect to $\Delta$,
and not to $\tau_c$.


Any quantum computer code is just a rotation in the Hilbert space of
the qubits and can be described as a trajectory on
$\mathbb{CP}^{2N-1}$, where $N$ is the total number of qubits. In the
Schr\"odinger picture, the evolution is given by the natural action on
$S^{4N-1}$ by $SU(2^{N})$. The most general fault-tolerant quantum
circuit is therefore defined by the Hamiltonian
%
$H_{\rm QC}(t) = \sum b_{j}(t)\, e_{j}$,
%
where $\left\{ e_{j} \right\}$ are the generators of the Lie algebra
of $SU(2^{N})$. The evolution operator associated with this
Hamiltonian satisfies the integral equation
\begin{eqnarray} 
W(t,0) & = & 1 - \frac{i}{\hbar} \int_{0}^{t} dt^{\prime}
H_{\rm QC}(t^{\prime})\, W(t^{\prime},0) \nonumber \\ & = & T_{t}\,
e^{-\frac{i}{\hbar} \int_{0}^{t} dt^{\prime} H_{\rm QC}(t^{\prime})},
\label{eq:Wdef}
\end{eqnarray}
such that the computer state vector at time $t$ is given by $\left|
\psi(t) \right\rangle = W(t,0) \left| \psi(0) \right\rangle$, where
$\left|\psi(0) \right\rangle$ represents the initial state of the
computer. Therefore, in the interaction picture, the interaction
operator is given by
\begin{eqnarray}
V(t) & = &  W^{\dagger}(t)\,e^{\frac{i}{\hbar} H_{0}t} \, V\,
e^{-\frac{i}{\hbar} H_{0}t} W(t) \\ & = & \sum_{{\mathbf{x}}}
\sum_{\alpha = \{ x,y,z\}} \frac{\lambda_{\alpha}}{2} \left[
e^{\frac{i}{\hbar} H_{0}t} f_{\alpha}({\mathbf{x}})\,
e^{-\frac{i}{\hbar}H_{0}t} \right] \nonumber \\ 
& & \qquad \times\,
W^{\dagger}(t)\, \sigma^{\alpha}\, ({\mathbf{x}}) W(t) \nonumber \\ 
& = & \sum_{{\mathbf{x}}} \sum_{\alpha = \{x,y,z\}}\!\!
\frac{\lambda_{\alpha}}{2} f_{\alpha} ({\mathbf{x}},t) W^{\dagger}(t)
\sigma^{\alpha} ({\mathbf{x}}) W(t).\quad
\end{eqnarray}
Since $W(t)$ is a $SU(2^{N})$ matrix, then
\begin{equation}
G^{\alpha} ({\mathbf{x}},t) = W^{\dagger}(t)\, \sigma^{\alpha}\,
({\mathbf{x}})W(t)\label{eq:GofT}
\end{equation}
is another matrix of $SU(2^{N})$, and we can write
\begin{equation} 
V(t) = \sum_{{\mathbf{x}}}\sum_{\alpha = \{
x,y,z\}}\frac{\lambda_{\alpha}}{2} f_{\alpha}({\mathbf{x}},t)\,
G^{\alpha}({\mathbf{x}},t).\label{eq:bigG}
\end{equation}

Although the expression in Eq.~(\ref{eq:bigG}) is general, it is not
very instructive. Furthermore, it is very undesirable from an error
correction standpoint: since $G(t)$ is an arbitrary matrix of
$SU(2^{N})$, the $V(t)$ in Eq.~(\ref{eq:bigG}) in principle generates
a highly complex correlated error that is nevertheless first order in
the coupling to the environment. The problem with the derivation of
Eq.~(\ref{eq:bigG}) is that it is too general since we assumed that
arbitrary rotations are performed at each single step. However, one of
the cornerstones of quantum computation is that such general rotations
can be approximately decomposed into a series of elementary gates
\cite{NC00}. Hence, our strategy will be to specialize the calculation
to these elementary gates and assume that general rotations can be
implemented by a finite series of such gates which are well resolved
in time.

\subsection{Single-qubit operations}
\label{sec:1qoperation}

When only single-qubit operations are performed, we have
\begin{equation} 
H_{\rm QC}\left(t\right) = \sum_{{\mathbf{x}}} \sum_{\alpha = \left\{
x,y,z\right\}} b_{\alpha} \left({\mathbf{x}},t\right)
\sigma^{\alpha}\left({\mathbf{x}}\right).
\end{equation}
In this case, $W(t)$ is the product of $SU(2)$ matrices
acting in each qubit's Hilbert space. Thus,
$G^{\alpha}({\mathbf{x}},t)$ simplifies to
\begin{equation}
G_1^{\alpha}({\mathbf{x}},t) = \left[\begin{array}{cc}
\rho_{1}e^{-i\phi} & -\rho_{2}e^{i\varphi}\\ \rho_{2}e^{-i\varphi} &
\rho_{1}e^{i\phi} \end{array} \!\right]\! \sigma^{\alpha}({\mathbf{x}})
\!\left[\! \begin{array}{cc} \rho_{1}e^{i\phi} & \rho_{2}e^{i\varphi}\\
-\rho_{2}e^{-i\varphi} & \rho_1 e^{-i\phi}\end{array}\!\right] ,
\label{eq:SU2rotation}
\end{equation}
where $\rho_{1}^{2}+\rho_{2}^{2}=1$ and $\left\{
\rho_{1},\rho_{2},\phi,\varphi\right\} $ are functions of
$\mathbf{x}$ and $t$. The single-qubit rotations yield an
expression of the form
\begin{equation} 
G_1^{\alpha} ({\mathbf{x}},t) = \sum_{\beta=\{ 1,x,y,z\}}
g^{\alpha\beta} ({\mathbf{x}},t) \, \sigma^{\beta} ({\mathbf{x}})
\label{eq:1-qubitgates}
\end{equation}
for some $g^{\alpha\beta} ({\mathbf{x}},t)$. By decomposing the
operators $f_{\alpha}$ and functions $g^{\alpha\beta}$ into their
Fourier components, we can give a more formal meaning to ``fast'' and
``slow'' gates,
\begin{equation}
f_{\alpha}({\mathbf{x}},t)\, g^{\alpha\beta}({\mathbf{x}},t) =
\!\sum_{\left|\omega_{1}\right|<\Lambda,\omega_{2}}\!\!\!
e^{i\left(\omega_{1}+\omega_{2}\right)t}
f_{\alpha}({\mathbf{x}},\omega_{1})\,
g^{\alpha\beta}({\mathbf{x}},\omega_{2}).
\end{equation}
Hence, if we define $\nu = \omega_{1}+\omega_{2}$, we can rewrite the
perturbation as
\begin{equation}
V = \sum_{\beta}\left\{ \sum_{\nu}e^{i\nu t} \left[ \sum_{\omega_{2}}
 \sum_{\alpha} f_{\alpha} \left( \nu - \omega_{2} \right)
 g^{\alpha\beta} \left( \omega_{2} \right) \right] \right\}
 \sigma_{\beta}.
\end{equation}
In the limit of fast gates, $|\omega_{2}|>\Lambda$, $f$ and $g$ are
not convolved, since they have distinct frequency domains. Therefore,
the noise operators $f_{\alpha}$ are unaltered by the
rotation. However, if $g$ has a significant weight at frequencies
smaller than $\Lambda$ (slow gates), one must convolve $f$ with $g$,
yielding a substantially different noise operator.

\subsection{Two-qubit operations}
\label{sec:2qoperation}

The general Hamiltonian for two-qubit gates is of the
form
\begin{equation}
H_{\rm QC}\left(t\right) = \sum_{{\mathbf{x}},{\mathbf{y}}}
\sum_{\alpha,\beta=\left\{ x,y,z\right\}} J^{\alpha\beta}
\left({\mathbf{x}},{\mathbf{y}},t\right)\,
\sigma^{\alpha}({\mathbf{x}})\, \sigma^{\beta}({\mathbf{y}}).
\end{equation}
However, one can also generate a full set of gates using instead a
single type of interaction,
%
\begin{equation} 
H_{\rm QC}\left(t\right) = \sum_{{\mathbf{x}},{\mathbf{y}}} J
\left({\mathbf{x}},{\mathbf{y}},t\right)\, \sigma^{a} ({\mathbf{x}})\,
\sigma^{b} ({\mathbf{y}}) \label{Ising}
\end{equation}
where $a$ and $b$ are fixed for each gate
$({\mathbf{x}},{\mathbf{y}})$. In order to see that this is sufficient
we can for instance set $a=b=z$. This generates the liquid NMR
Hamiltonian \cite{VAN02}, where the Ising interaction,
Eq.~(\ref{Ising}), and single qubit rotations can be used to generate
a control $\sigma^z$ gate.

We keep $a$ and $b$ arbitrary. However, for the sake of simplicity, we
assume that only operations between disjoint pairs are allowed; that
is, if $J\left({\mathbf{x}},{\mathbf{y}}_{1},t\right) \neq 0$, then
$J\left({\mathbf{x}},{\mathbf{y}}_{2},t\right) = 0$ for all
${\mathbf{y}}_{2} \neq {\mathbf{y}}_{1}$. It is then straightforward
to write down $W(t)$ in a compact form: The time-ordering
[Eq.~(\ref{eq:Wdef})] is automatically taken care of by the sequence
of gates, while for a gate involving qubits ${\mathbf{x}}$ and
${\mathbf{y}}$ the contribution to $W(t)$ is
\begin{eqnarray}
W \left({\mathbf{x}},{\mathbf{y}},t\right) & = & \cos \left[
\theta({\mathbf{x}},{\mathbf{y}},t) \right] \\ & & +\; i \sin\left[
\theta({\mathbf{x}},{\mathbf{y}},t) \right] \nonumber
\sigma^{a}({\mathbf{x}})\, \sigma^{b}({\mathbf{y}}),
\label{eq:rotation2qubits}
\end{eqnarray}
where $\theta({\mathbf{x}},{\mathbf{y}},t) = \int_{0}^{t}
dt^{\prime} J({\mathbf{x}},{\mathbf{y}},t^{\prime})$.
Hence, a two-qubit rotation yields
\begin{eqnarray}
G_2^{\alpha}({\mathbf{x}},t) & = & \sin \left[
 2\theta({\mathbf{x}},{\mathbf{y}},t) \right]
 \epsilon^{a\alpha\gamma} \sigma^{\gamma}({\mathbf{x}})
 \sigma^{b}({\mathbf{y}})\nonumber \\ & + & \cos \left[
 2\theta({\mathbf{x}},{\mathbf{y}},t) \right] \left(
 1-\delta_{a,\alpha} \right) \sigma^{\alpha}({\mathbf{x}})\nonumber \\
 & + & \delta_{a,\alpha}\sigma^{\alpha}({\mathbf{x}}),
\label{eq:2-qubitsgates}
\end{eqnarray}
where $\epsilon^{a\alpha\gamma}$ is the usual antisymmetric tensor.

The first term on the \emph{r.h.s.}\ of Eq.~(\ref{eq:2-qubitsgates})
tells us that the 2-qubit gate can propagate the error from the qubit
at $\mathbf{x}$ to the qubit at position $\mathbf{y}$.  However, it
also tells us that it is possible to choose a particular gate where
this propagation does not happen (by choosing $a\!=\!\alpha$, for
instance).  Unfortunately, propagating errors in the quantum circuit
is in general unavoidable (since the only gate that commutes with all
Pauli operators is the identity).

The second and third terms on the \emph{r.h.s.} of
Eq.~(\ref{eq:2-qubitsgates}) are much less dramatic. They simply
describe a local noise that is not propagated by the gate.

\subsection{Upper-bounds for the evolution}
\label{sec:upperbounds}

In Eqs. (\ref{eq:1-qubitgates}) and (\ref{eq:2-qubitsgates}), we
showed that one- and two-qubit gates can introduce what is seemingly a
very complicated noise structure. The expressions depend on how the
gates are implemented, thus hiding a general assessment. We can
advance the discussion by recalling that $W$ is always an unitary
matrix. Hence, the coefficients in Eqs.(\ref{eq:1-qubitgates}) and
(\ref{eq:2-qubitsgates}) have modulus equal or smaller than unity. A
suitable upper bound on the effects of slow gates is then provided by
setting all these coefficients equal to one. Thus, the operators
expressed in Eqs. (\ref{eq:1-qubitgates}) and (\ref{eq:2-qubitsgates})
gain the upper bounds
\begin{eqnarray}
\tilde{G}_1^{\alpha}({\mathbf{x}}) & = & \sum_{\beta=\{x,y,z\}}
\sigma^{\beta} ({\mathbf{x}}), \\ \tilde{G}_2^{\alpha}({\mathbf{x}}) &
= & \sigma^{\alpha} ({\mathbf{x}}) + \epsilon^{a\alpha\gamma}
\sigma^{\gamma}(\mathbf{x})\sigma^{b}(\mathbf{y}) \label{G2-b}.
\end{eqnarray}

$\tilde{G}_2^{\alpha}$ still looks troublesome, since it tells us that
an error in qubit $\mathbf{x}$ is propagated to $\mathbf{y}$. However,
this is not a problem of the finite gate time operation, since an
instantaneous and perfect gate will also propagate the error in a
similar fashion. In order to obtain an upper bound for the effects
introduced by the two-qubit gates, we precisely follow this fact. We
consider that all the qubit components are exposed to all the noise
channels all the time. Thus, we replace Eq.~(\ref{G2-b}) by
$G_2^{\alpha}({\mathbf{x}}) = \sum_{\beta=\{x,y,z\}}
\sigma^{\beta}({\mathbf{x}})$ and assume that two-qubit gates are
performed instantaneously. In summary, we reduce the problem of finite
time operation of the two-qubit gate to the problem of a noisier qubit
environment and propagating errors in the quantum code by perfect
gates. Now we can rely on the theory of fault-tolerance
\cite{NC00,AGP06}, and simply assume that the error propagation is
handled by the quantum code.

The final conclusion is that an upper bound estimate on the effects of
slow gates is obtained by the interaction Hamiltonian
\begin{equation}
V_{\rm eff}(t) = \sum_{{\mathbf{x}}} \sum_{\alpha=\{x,y,z\}}
\frac{\lambda}{2} f_{\rm eff}({\mathbf{x}},t) \,
\sigma^{\alpha}({\mathbf{x}},t),
\label{eq:bigF}
\end{equation}
where
\begin{equation}
f_{\rm eff}\left({\mathbf{x}},t\right) = \frac{1}{\lambda^{2}} \left[
\sum_{\beta=\left\{ x,y,z\right\}} \lambda_{\beta} f_{\beta}
\left({\mathbf{x}},t\right) \right]
\end{equation}
and $\lambda = \sqrt{\sum_{\beta=\left\{ x,y,z\right\}}
\lambda_{\beta}^{2}}$ is the new coupling parameter. Although this is
a brutal approximation, it will be sufficient for our discussion. As
we will argue later, for the purpose of determining the effect of
long-wavelength correlations on the threshold theorem, the only
relevant aspect of the $f_{\alpha}$ is their scaling dimension. Since
$\dim f_{\rm eff}$ is in general equal to $\min\left(\dim
f_{\alpha}\right)$, it is sufficient to use Eq.~(\ref{eq:bigF}) as the
worst case scenario.

\emph{Thus, in both limiting cases, fast and slow gates, we arrive at
the same functional form for the effective interaction.} Hence, both
cases can be handled simultaneously, and we proceed to the analysis of
QEC in the presence of this interaction. In order to simplify the
notation, we hereafter drop the subscript ``eff'' from the slow-gate
operators.

\section{Quantum Error Correction in Correlated~Environments}
\label{sub:QEC-and-perturbation}

A QEC code is defined as the combination of encoding, decoding, and
recovery operations. Since we were able to make our analysis code
independent, the unitary component of the QEC protocol is described by
$U(\Delta,0)$, Eq.~(\ref{eq:unitaryevolution}), with the appropriate
$V(t)$ discussed in Sec. \ref{sec:perturbation}. The final ingredient
in standard QEC is just the syndrome extraction $P$, which is a
projective measurement.

In Ref.~\onlinecite{NB06} it was demonstrated how to define $P$ and
its effects on $U$ for stabilizer error correction codes. It is
important to remark that an error which keeps the computer in the
logical Hilbert space can never be corrected by QEC. This is simply a
statement that for the general assumptions we make, the problem of
protecting quantum information never satisfies the second criterion of
Lafflame-Knill \cite{KL97} for perfect QEC.  In simple terms, the
criteria states that all allowed errors must always take the logical
one and logical zero to orthogonal states [Eq.~(20) of
Ref.~\onlinecite{KL97}]. By construction, these errors are high-order
events in the coupling with the environment. Nevertheless, as we
already know (\ref{sec:the-problem-of}), this fact per se is not
enough to ensure that such errors will not be relevant at long
times. One of our goals is to find out when it is appropriate to
safely neglect such uncorrectable errors in the presence of correlated
environments.

\begin{figure}[b]
\subfigure[]{\includegraphics[width=4cm]{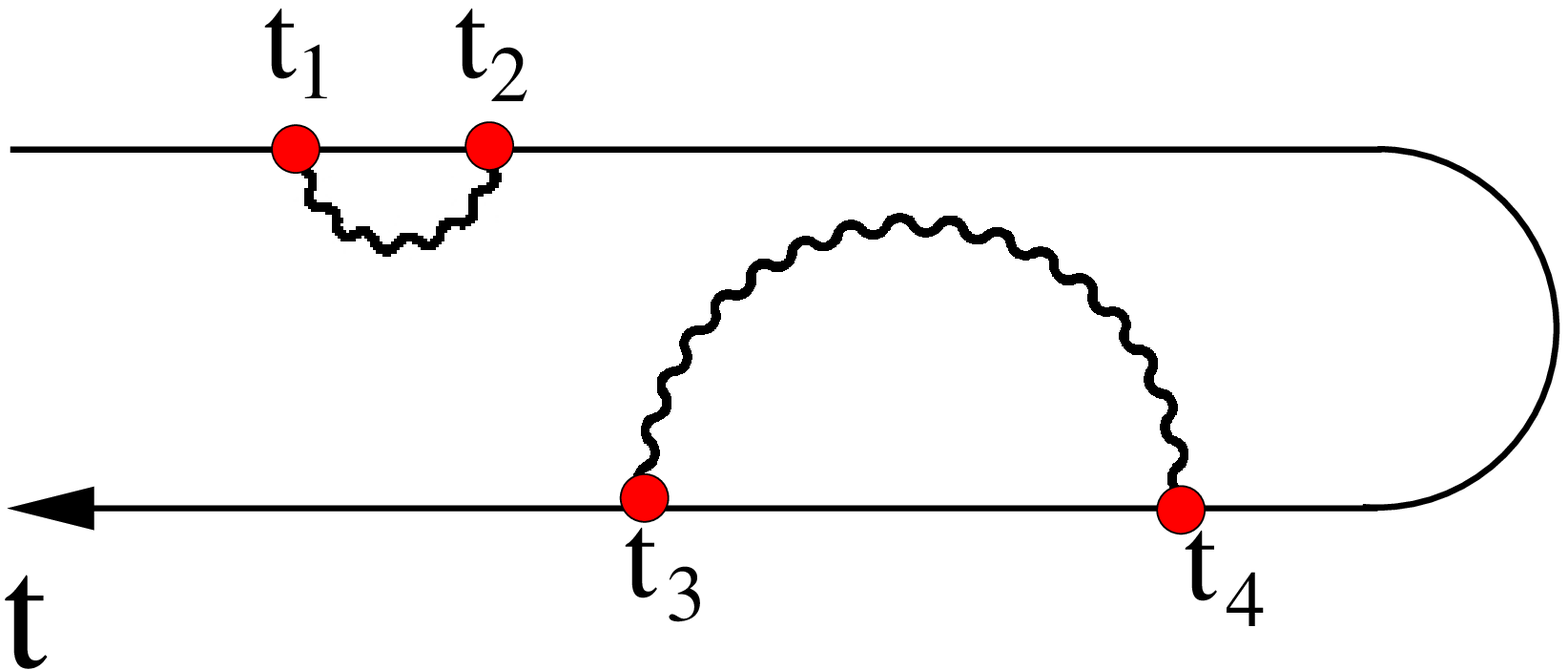}}
\subfigure[]{\includegraphics[width=4cm]{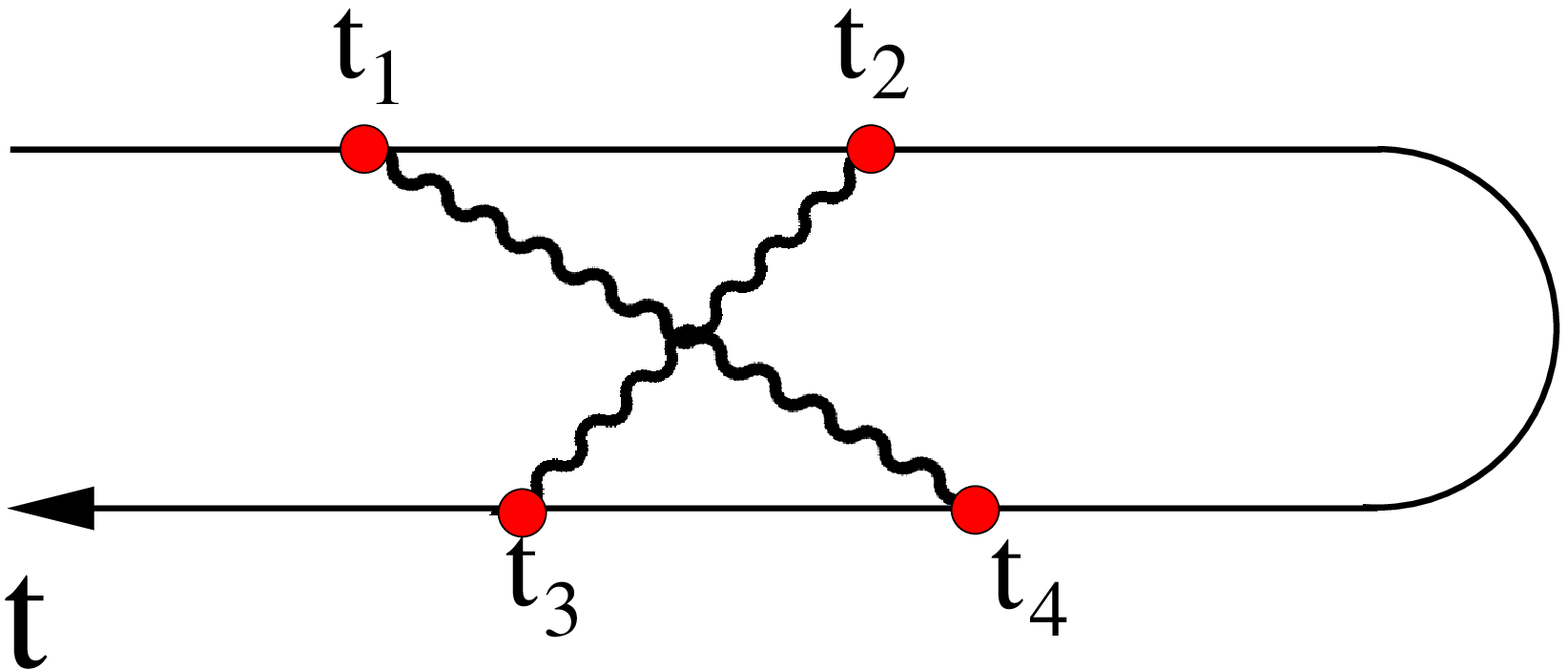}}
\subfigure[]{\includegraphics[width=4cm]{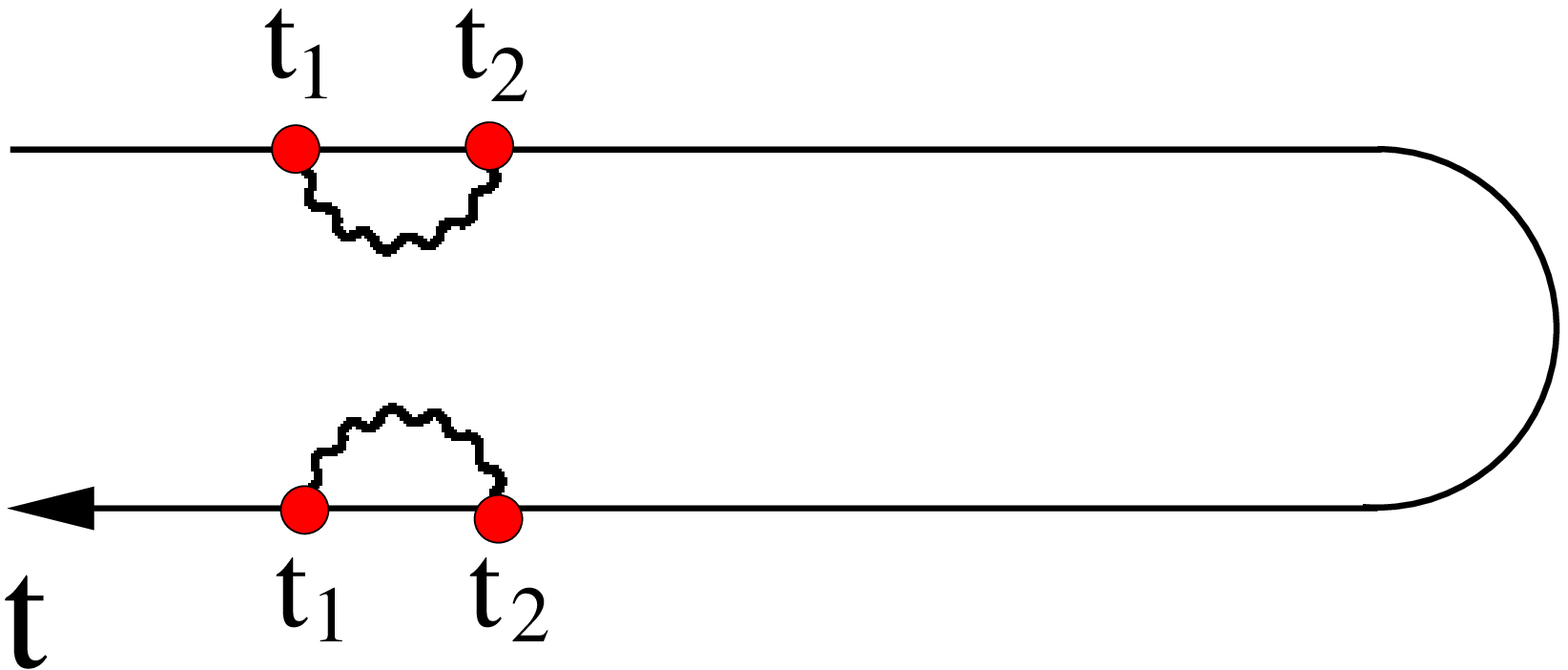}}
\subfigure[]{\includegraphics[width=4cm]{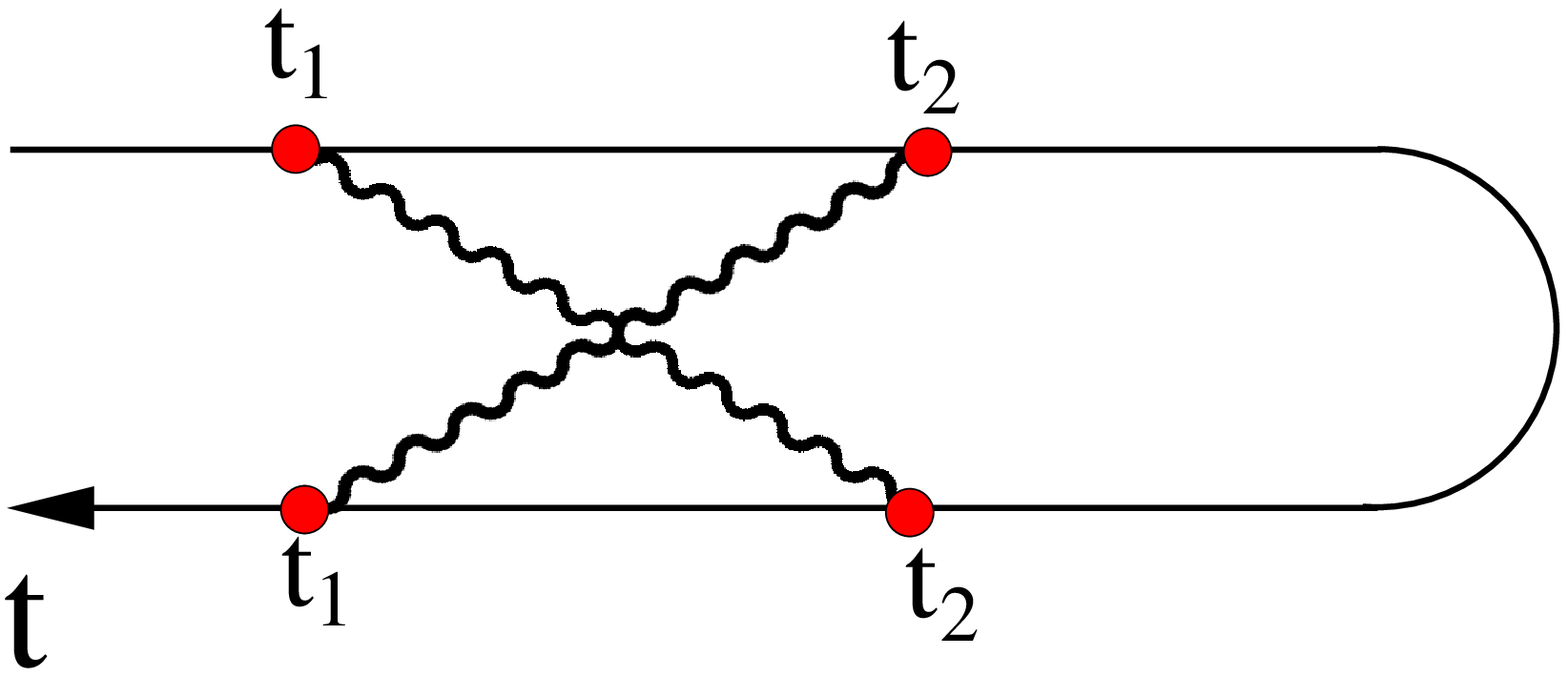}}
\caption{Graphical representation of a few fourth-order terms in a
``time-loop'' expansion for either the probability of a given
evolution or the reduced density matrix (spatial dimensions are
suppressed for clarity). Points of interaction with the bath (circles)
are connected by propagation of the environmental modes (wiggly
lines). In the top diagrams, the time integrals are unconstrained, as
would be the case for unitary evolution. In the bottom diagrams, the
detection of an error by a QEC protocol forces the interactions with
the bath to occur at the same times on both the forward and backward
legs in order that $U$ and $U^\dagger$ correspond to the same
syndrome. This additional constraint introduced by QEC is crucial in
the long-time behavior.  }
\label{fig:Keldysh}
\end{figure}

In hindsight, it is not hard to understand the benefits of QEC. Thus,
for the sake of readability, we present first a qualitative argument
that captures the overall discussion.

As we defined in the introduction, there are two quantities that we
are interested in calculating: (i) the probability of a given
evolution, and (ii) the reduced density matrix of the computer.  Both
quantities are written as a double series in the coupling with the
environment. On the one hand, the initial ket of computer and the
environment, $|\Psi\rangle$, evolves in the time interval $[0,t]$ by
the time ordered series $U(t)$. On the other hand, the bra
$\langle\Psi|$ evolves in time with the anti-time-ordered series
$U^\dagger$. It is only a subset of each series that enters in the
evaluation of either the probability or the reduced density matrix,
because of the measurements present in the traditional formulation of
QEC. Hence, it is usually a non-trivial task to calculate the
necessary expectation values.

Because we are dealing with a double series, it is natural to use a
formalism analogous to a time-loop expansion \cite{mahan}. There are
six (interrelated) Green functions in such an expansion: The usual
advanced and retarded functions for the time-ordered series; the
advanced and retarded functions for the anti-time-ordered series; and
the lesser and greater functions, which contract a term from the
time-ordered series with another one from the anti-time-ordered
series. This formalism is often referred to as the Schwinger-Keldysh
approach \cite{Sch61,Kel65}. It is usually represented graphically by
a double contour in time (see Fig.~\ref{fig:Keldysh}). The upper leg
stands for the time-ordered evolution for the time interval $[0,t]$,
while the lower leg stands for the anti-time-ordered evolution in the
reversed interval $[t,0]$.

Let us for the moment assume that a short-time expansion is valid and
focus on a single qubit. Then, the evolution operator for
that particular qubit within a QEC cycle is given by
\begin{widetext}
\begin{eqnarray}
U_1\left(\Delta,0\right) & \approx & 1 - \frac{i}{\hbar}
\int_{0}^{\Delta} dt \sum_{\alpha=\{x,y,z\}}
\frac{\lambda_{\alpha}}{2} f_{\alpha}({\mathbf{x}},t)\,
\sigma^{\alpha}({\mathbf{x}},t) \nonumber \\ & & -\,
\frac{1}{\hbar^{2}} \int_{0}^{\Delta} dt \int_{0}^{t} dt^{\prime}
\sum_{\alpha=\{x,y,z\}} \frac{\lambda_{\alpha}\lambda_{\beta}}{4}
f_{\alpha}({\mathbf{x}},t) f_{\beta}({\mathbf{x}},t^{\prime})\,
\sigma^{\alpha}({\mathbf{x}},t) \sigma^{\beta}
({\mathbf{x}},t^{\prime})  +\ O(\lambda^3).
\label{eq:lowestorderU}
\end{eqnarray}
\end{widetext}
In Fig.~\ref{fig:Keldysh} we represent graphically a few terms of
order $\lambda^{4}$.  All of these terms are the product of a
second-order term from $U_1$ and a second-order term from
$U_1^\dagger$ [see Eq.~(\ref{eq:lowestorderU})]. Hence, they
correspond to two ``errors'' in the qubit evolution and involve the
expectation value
\begin{equation}
 \left\langle \Psi \right| f_{\alpha}^{\dagger}({\mathbf{x}},t)
f_{\beta}^{\dagger}({\mathbf{x}},t^{\prime})
f_{\alpha}({\mathbf{x}},t^{\prime \prime})
f_{\beta}({\mathbf{x}},t^{\prime \prime \prime}) \left| \Psi
\right\rangle . 
\label{expectationV}
\end{equation}
Using Wick's theorem, we can immediately write (\ref{expectationV}) as
a product of the non-interacting Green functions. Each possible set of
contractions leads to the different ``diagrams'' in
Fig.~\ref{fig:Keldysh}.

We usually do not know when an ``error'' occurs; hence, each Green
function is accompanied in the series by a double integral in time.
This is precisely the case in an unprotected computer's evolution or
inside a QEC cycle [see Figs.\ref{fig:Keldysh} (a) and (b)].  However,
a dramatic change happens in a Green function between terms for
different cycles. When the syndrome shows that a particular error
occurred in a certain QEC cycle, we can re-write
Eq. (\ref{expectationV}) to reflect this knowledge:
\begin{equation}
 \left\langle \Psi \right| f_{\alpha}^{\dagger}({\mathbf{x}},t)
f_{\beta}^{\dagger}({\mathbf{x}},t^{\prime})
f_{\alpha}({\mathbf{x}},t+\delta t)
f_{\beta}({\mathbf{x}},t^{\prime}+\delta t^{\prime}) \left| \Psi
\right\rangle , \label{expectationV2}
\end{equation}
where $\delta t$ and $\delta t^{\prime}$ are time variables with range
smaller than the QEC period.  After integrating the ``high frequency''
part (the $\delta t$ and $\delta t^{\prime}$ variables), we end up
reducing Eq. (\ref{expectationV2}) to
\begin{equation}
 \left\langle \Psi \right| f_{\alpha}^{\dagger}({\mathbf{x}},t)
f_{\beta}^{\dagger}({\mathbf{x}},t^{\prime})
f_{\alpha}({\mathbf{x}},t) f_{\beta}({\mathbf{x}},t^{\prime}) \left|
\Psi \right\rangle  \label{expectationV3}
\end{equation}
with $t$ and $t^\prime$ representing a coarse-grained time scale of
order the QEC period [see Figs. \ref{fig:Keldysh} (c) and (d)].
Therefore, although we are considering terms of the same order in
$\lambda$, the number of ``time integrals'' in the coarse-grained
scale (low frequencies) is half that in the original microscopic
calculation (high frequencies).

The simple dimensional analysis of Sec.~\ref{sec:the-problem-of}
tells us now that QEC has changed the criteria for the stability of the
perturbation series at long times. As we demonstrate now,
it is less stringent than the naive expectation.

\subsection{Quantum evolution steered by QEC}

It is reasonable to assume that at the beginning of the computation
the computer's state vector, $\psi_{0}$, and the environment's,
$\varphi_{0}$, are not entangled,
\begin{equation}
\left| \Psi \left(t=0\right) \right\rangle = \left| \psi_{0}
\right\rangle \otimes \left|\varphi_{0} \right\rangle.
\end{equation}
In a realistic situation, $\psi_0$ would have some initialization
error and be entangled with the environment to some degree (both of
which would yield errors in $\psi$). However, here we neglect these
effects in order to keep the discussion focused.

Just as in the case of the 3-qubit code, by the end of a QEC cycle the
computer will have evolved according to the unitary operator
$U(\Delta,0)$. Then, the syndrome is extracted and the computer wave
function is projected,
\begin{eqnarray}
& & P_{m}\, U \left(\Delta,0\right) \left| \Psi \left(0\right)
\right\rangle,
\end{eqnarray}
where $m$ corresponds to a particular syndrome, with $\sum_{m}P_{m}
\!=\! I$ and $P_{m}^{2} \!=\!  P_{m}$. In the case of many logical
qubits evolving together, then $m$ denotes the set of all the
syndromes extracted at time $\Delta$. The last step in the code is the
appropriate recovery operation, $R_{m}$, depending on the syndrome
outcome,
\begin{equation}
\left| \Psi \left(\Delta\right) \right\rangle = R_{m}
\left(\Delta+\delta,\Delta\right)\, P_{m}\, U
\left(\Delta,0\right) \left| \Psi \left(0\right) \right\rangle.
\end{equation}

Since in a fault-tolerant error correction scheme the information is
never decoded (in contrast to the 3-qubit code discussed above), the
quantum information always remains protected. Therefore, we can deal
with our two limiting cases (slow- and fast-gates) in two different
ways. In the case of a slow-gate recovery, we formally include it as
the initial step of the next QEC period. Conversely, in the case of
fast gates, we assume that the recovery is performed flawlessly in a
very short time scale after the projection. For the sake of clarity,
we choose the latter below. We emphasize that this does not restrict
our discussion, since it is known that the time of recovery is
irrelevant to the error correction. In fact, it can be postponed all
the way to the end of the calculation \cite{DA07}.

\subsection{Probability of a syndrome history and the loss 
of information}
\label{sec:faultpath}

The first quantity to discuss is the probability of
measuring a particular syndrome at the end of the first QEC step,
\begin{equation}
\mathcal{P}_m = \left\langle \Psi (0) \right| U^{\dagger}
(\Delta,0)\, P_m\, U (\Delta,0) \left| \Psi(0) \right\rangle.
\label{eq:localp}
\end{equation}
The corresponding reduced density matrix is
\begin{widetext}
\begin{equation}
\rho_{\vec{r},\vec{s}}^m (\Delta) =
\frac{\mbox{tr}_{\varepsilon} \left[ \left\langle \vec{r} \right|
P_m U (\Delta,0) \left|
\Psi (0) \right\rangle \left\langle \Psi (0) \right| U^{\dagger}
(\Delta,0) P_m\left| \vec{s} \right\rangle
\right]} {\left\langle \Psi (0) \right|
U^{\dagger} (\Delta,0) P_{m} U (\Delta,0) \left| \Psi (0)
\right\rangle} = \frac{\left\langle \varphi_{0} \right| \left[
\left\langle \psi_{0} \right| U^{\dagger} (\Delta,0) P_m
\left| \vec{s} \right\rangle \left\langle \vec{r} \right| P_m
U (\Delta,0) \left| \psi_{0} \right\rangle \right] \left| \varphi_{0}
\right\rangle}{\left\langle \varphi_{0} \right| \left\langle \psi_{0}
\right| U^{\dagger} (\Delta,0) P_{m} U (\Delta,0) \left|
\psi_{0} \right\rangle \left|\varphi_{0} \right\rangle},
\label{eq:localdensity}
\end{equation}
\end{widetext}
where $\vec{r}$ and $\vec{s}$ denote states in the computer Hilbert
space and $\mbox{tr}_{\varepsilon}$ is the trace over the environment
Hilbert space. It is possible to quantify how much information was
leaked to the environment by calculating the von Neumann entropy
\begin{equation}
S\left(\Delta\right) = - \mbox{tr}_{c}\, \left[ \rho^m
(\Delta) \ln \left| \rho^m (\Delta) \right| \right],
\end{equation}
where $\mbox{tr}_{c}$ is the trace over the computer Hilbert space.

In Eqs.~(\ref{eq:localp}) and (\ref{eq:localdensity}), one clearly
sees the important role played by the projection operators in the
quantum evolution steered by QEC. The careful construction of the
encoded states combined with the measurement (syndromes) reduces the
quantum interference between different history paths of the
computer. By {\it partially} collapsing the wave function of the
computer, this traditional form of QEC reduces decoherence.

Equations (\ref{eq:localp}) and (\ref{eq:localdensity}) define the
local components of the noise. When spatial correlation between qubits
can be ignored, they are related to the stochastic probabilities and
density matrix discussed in Sec.~\ref{sec:QEC} [see Eqs. (\ref{eq:Pm})
and (\ref{eq:rhoud})].

The generalization to a sequence of QEC cycles is straightforward
\cite{NB06},
\begin{equation}
\Upsilon_{{\mathbf{w}}} = \upsilon_{w_{N}} \big( N \Delta, (N-1)
\Delta \big) ... \upsilon_{w_{1}} (\Delta,0),
\label{eq:defUpsilon}
\end{equation}
where ${\mathbf{w}}$ is the particular history of syndromes for all
the qubits and
%
%
\begin{eqnarray}
\lefteqn{\upsilon_{w_{j}} \big(j\Delta,(j-1) \Delta \big) =} & & \nonumber\\
& &  R_{w_{j}} \big( j
(\Delta+\delta), j \Delta \big) P_{w_{j}} U \big( j \Delta, (j-1)
\Delta \big),
\label{eq:v-evolution}
\end{eqnarray}
%
%
is the QEC evolution after each cycle. Each history comes with the
associated probability
\begin{equation}
\mathcal{P} \left( \Upsilon_{{\mathbf{w}}} \right) = \left\langle
\varphi_{0} \right| \left\langle \psi_{0} \right|
\Upsilon_{{\mathbf{w}}}^{\dagger} \Upsilon_{{\mathbf{w}}} \left|
\psi_{0} \right\rangle \left| \varphi_{0} \right\rangle.
\label{eq:hisprob}
\end{equation}
Finally, there is always some residual decoherence which can be found
from the reduced density matrix
\begin{equation}
\rho_{\vec{r},\vec{s}} \left( \Upsilon_{{\mathbf{w}}} \right) =
\frac{\left\langle \varphi_{0} \right| \left[ \left\langle \psi_{0}
\right| \Upsilon_{{\mathbf{w}}}^{\dagger} \left| \vec{s} \right\rangle
\left\langle \vec{r} \right| \Upsilon_{{\mathbf{w}}} \left| \psi_{0}
\right\rangle \right]\left|\varphi_{0}\right\rangle }{\left\langle
\varphi_{0} \right| \left\langle \psi_{0} \right|
\Upsilon_{{\mathbf{w}}}^{\dagger} \Upsilon_{{\mathbf{w}}} \left|
\psi_{0} \right\rangle \left| \varphi_{0} \right\rangle},
\label{eq:rdmlqubit}
\end{equation}
with $\vec{r}$ and $\vec{s}$ being elements of the logical
subspace. This in turn yields the entropy
\begin{equation} 
S \left(\Upsilon_{{\mathbf{w}}}\right) = - \mbox{tr}_{c} \big[ \rho
\left(\Upsilon_{{\mathbf{w}}} \right) \ln \left| \rho
\left(\Upsilon_{{\mathbf{w}}} \right) \right| \big] \;.
\end{equation}

In the following, we will show for Eqs.  (\ref{eq:hisprob}) and
(\ref{eq:rdmlqubit}) how to separate the effect of correlations
between different QEC cycles from the contributions due to the local
component of the noise, as defined by Eqs. (\ref{eq:localp}) and
(\ref{eq:localdensity}).

\begin{figure}[b]
\includegraphics[width=5cm]{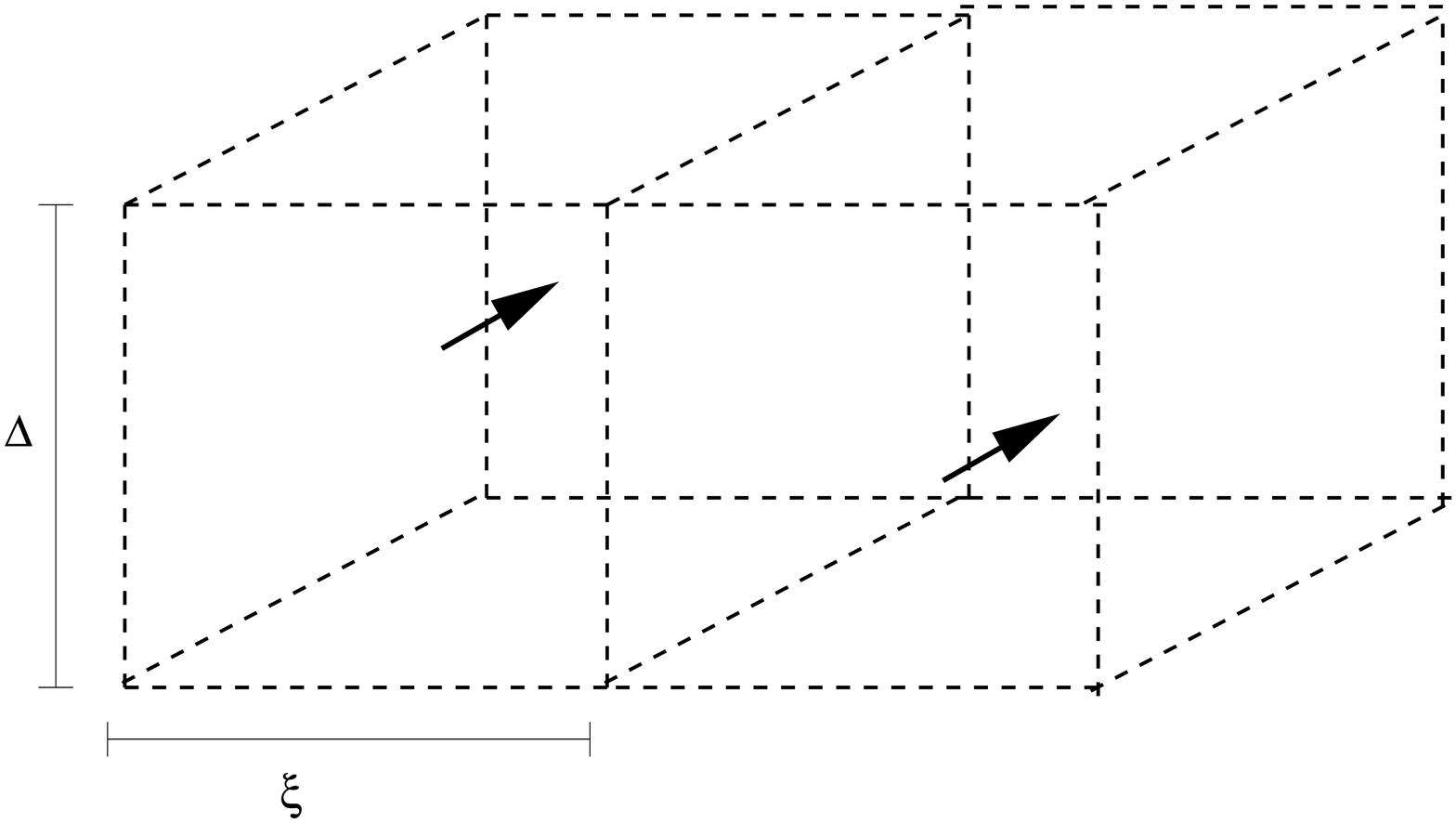}
\caption{Two neighboring hypercubes in space-time, each one containing
a qubit.}
\label{fig:hypercubes}
\end{figure}

\section{Perturbation theory and the hypercube~assumption}
\label{sec:hypercube}

There is one additional issue that we must deal with before we can
move forward. In principle, even the first order term in
Eq.~(\ref{eq:lowestorderU}) is already beyond the QEC approach that
has been outlined so far. The reason is that when calculating
$\mathcal{P}$ or $\rho$ we generate pair contractions of the type
$\left\langle f_{\alpha}({\mathbf{x}},t)
f_{\alpha}({\mathbf{y}},t^{\prime}) \right\rangle$. Therefore, the
probability of finding an error at a given qubit is conditional on
what happens with all other qubits. This automatically hinders the
simple probabilistic interpretation of QEC that we used in
Sec.~\ref{sec:QEC}.

The fact that we do not want to deal with such conditional
probabilities leads us to the single most important simplifying
hypothesis of our work: We assume that the qubits are separated by a
minimum distance
\begin{equation}
\xi=\left(\mathtt{v}\Delta\right)^{1/z},
\label{xi-def}
\end{equation}
where $\mathtt{v}$ is the excitation velocity and $z$ is the
dynamical exponent of the theory describing the environment. Hence,
for all ${\mathbf{x}} \neq {\mathbf{y}}$ and $\left| t-t^{\prime}
\right| < \Delta$, we have $\left\langle f_{\alpha}({\mathbf{x}},t)
f_{\alpha}({\mathbf{y}},t^{\prime}) \right\rangle \approx0$. It is
then possible to assign a probability for the short-time evolution of
each qubit independently of all others.

To further organize the analysis we order the qubits in a
$D$-dimensional array that defines hypercubes of volume $\Delta \times
\xi^{D}$ (see Fig.\ \ref{fig:hypercubes}). In summary, for times smaller than $\Delta$, each qubit has
a dynamics independent from the other qubits, hence resembling a
quantum impurity problem. However, for time scales larger than
$\Delta$, spatial correlations among them are present, thus making the
problem similar to a spin lattice.

Ideally, we would like to decompose the evolution operator in inter-
and intra-hypercube components,
\begin{equation}
U (\Delta,0) = U_{<} (\Delta,0)\, U_{>} (\Delta,0),
\end{equation}
where $<$ labels frequencies smaller than $\Delta^{-1}$ and $>$ frequencies in
the interval $\left[\Delta^{-1},\Lambda\right]$. Whenever this is possible, we
can integrate the intra-hypercube part in order to define a ``local'' evolution
and, consequently, a local error probability. There are simple noise models
where this can be done exactly \cite{NB06}, however, in general, this separation
is only possible in a perturbative expansion. Keeping just a few terms in
perturbation theory is not always adequate, and we must try to find ways to
improve it.

\subsection{Perturbation theory improved by RG}

Our objective in this section is to define an effective evolution operator that
can reasonably describe the evolution of the qubit within each hypercube. All 
terms consistent with the same syndrome and having the same leading long-time 
properties should be included. Within a hypercube, the environment induces interaction of a qubit only with itself; communication between qubits at longer times is treated in the next section. 

We use the renormalization group (RG) \cite{SHA94} to sum the most relevant 
families of terms in the perturbation series.
In order to improve the lowest order terms in the perturbation theory
through RG, we need to introduce the next higher-order terms in the
perturbation series. However, as we discussed previously, we are not
interested in the full unitary evolution, but rather the projected
terms obtained after the extraction of the syndrome. Therefore, in
order to apply RG to the first-order term, we need to consider
\begin{widetext}
\begin{eqnarray}
\upsilon_{\alpha}\left({\bf x}_{1},\lambda_{\alpha}\right) & \approx &
- \frac{i}{2\hbar} \lambda_{\alpha}\int_{0}^{\Delta}dt\,
f_{\alpha}\left({\bf x_{1}},t\right) - \frac{1}{8 \hbar^2} \left|
\epsilon_{\alpha\beta\gamma} \right| \lambda_{\beta}
\lambda_{\gamma}\, \sigma_{\alpha} \left(\Delta\right) T_{t}
\int_{0}^{\Delta} dt_{1}\, dt_{2}\, f_{\beta}\left({\bf x_{1}},t_{1}
\right) f_{\gamma}\left({\bf x_{1}},t_{2}\right) \sigma_{\beta}
\left(t_{1}\right) \sigma_{\gamma} \left(t_{2}\right)\nonumber \\ & &
+\, \frac{i}{48 \hbar^3} \sum_{{\bf \beta}} \lambda_{\alpha}
\lambda_{\beta}^{2}\, \sigma_{\alpha} \left(\Delta\right) T_{t}
\int_{0}^{\Delta} dt_{1}\, dt_{2}\, dt_{3}\, f_{\alpha} \left({\bf
x_{1}},t_{1} \right) f_{\beta} \left({\bf x_{1}},t_{2}\right)
f_{\beta} \left({\bf x_{1}},t_{3}\right) \sigma_{\alpha}
\left(t_{1}\right) \sigma_{\beta} \left(t_{2}\right) \sigma_{\beta}
\left(t_{3}\right),
\label{eq:extended-evolution}
\end{eqnarray}
\end{widetext}
where $\epsilon_{\alpha\beta\gamma}$ is the antisymmetric tensor
\cite{QubitTimeArg}. There is only one spatial index in 
(\ref{eq:extended-evolution}) because of the hypercube assumption: 
we have included only terms in which contraction of the $f$'s yields 
a non-zero value, as these will contribute to the effective short time
evolution. At long times, connections between the qubits are, of course, 
essential, and this is treated in the next section.

The RG is naturally implemented in the case of
ohmic baths (which leads to logarithmic singularities). However,
suitable generalizations can be defined by dimensional regularization
or by summing series in the expansion. Thus, in general, it is
possible to write the following beta function for $\upsilon_{{\bf
x}_{1}}^{\alpha}$:
\begin{equation} 
\frac{d\lambda_{\alpha}}{d\ell} = g_{\beta\gamma} \left(\ell\right)
\lambda_{\beta} \lambda_{\gamma} + \sum_{\beta} h_{\alpha\beta}
\left(\ell\right) \lambda_{\alpha} \lambda_{\beta}^{2},
\end{equation}
where $g$ and $h$ are functions specific to a particular environment,
$\ell \!=\! \Lambda/\Lambda^{\prime}$, and $\Lambda^{\prime}$ is the
reduced (i.e.\ rescaled) cutoff frequency. By integrating the beta
function from the bare cutoff, $\Lambda$, to $\Delta^{-1}$, we are
summing the most relevant components of the noise inside a
hypercube. If the renormalized value of the running coupling at
frequency $\Delta^{-1}$, $\lambda^\ast$, is still a small number, then
it is a good approximation to consider
\begin{equation}
\upsilon_{\alpha} \left({\bf x}_{1},\lambda_{\alpha}^\ast \right)
\approx \frac{-i\lambda_{\alpha}^\ast}{2\hbar} \int_{0}^{\Delta} dt\,
f_{\alpha} \left({\bf x}_{1},t\right)
\label{eq:v-evolution-reno}
\end{equation}
as the evolution operator of the qubit at position ${\bf x}_{1}$ which
was diagnosed with an error $\alpha$ by the QEC procedure.

We illustrate the renormalization group procedure with two simple
examples of ohmic baths: (i) marginally relevant and (ii) marginally
irrelevant couplings.

\subsubsection{The k-channel Kondo problem}

The first example is a qubit exposed to a bosonic bath that is modeled
by a $SU(2)_{k}$ Kac-Moody algebra -- the bosonized
Hamiltonian of a $k$-channel Kondo problem. Here we closely follow
the work of Affleck and Ludwig (see appendix B of
Ref.\,\onlinecite{AL91}). We define chiral bosonic currents
$:\vec{J_{L}}:$ obeying the operator product expansion (OPE)
\begin{equation}
:\!J_{L}^{a} \left(t \right)\!:\; :\!J_{L}^{b} \left( t^{\prime}
\right)\!: \to \frac{f^{abc}:\!J_{L}^{c}(t)\!:} {\mathtt{v} \left( t -
t^{\prime} \right)} - \frac{k\delta^{ab}} {2 \mathtt{v}^{2} \left( t -
t^{\prime} \right)^{2}},
\end{equation}
where $f^{abc}$ are the group structure constants and $\mathtt{v}$ is
the velocity of excitations. In the interaction picture, the qubit
couples to the currents by the usual Kondo interaction, yielding an
evolution operator (or, equivalently, a scattering matrix) of the form
\begin{equation}
U = T_{t}\, e^{-\frac{i\lambda \mathtt{v}} {2\hbar}
\int_{-\infty}^{\infty} dt\, :\vec{J}_{L} (t)\!: \cdot \vec{\sigma}}.
\label{eq:Skondo}
\end{equation}
Following our general discussion, we expand the evolution operator to
lowest order in the coupling,
\begin{widetext}
\begin{eqnarray}
U & \approx & 1 - \frac{i\lambda \mathtt{v}} {2\hbar}
 \int_{-\infty}^{\infty} dt :\vec{J}_{L}(t)\!: \cdot\, \vec{\sigma} -
 \left(\frac{\lambda \mathtt{v}} {2\hbar} \right)^{2} \sum_{a,b}
 \int_{-\infty}^{\infty} dt \int_{-\infty}^{t} dt^{\prime}
 :\!J_{L}^{a} (t)\!:\; :\!J_{L}^{b} \left(t^{\prime}\right)\!:
 \sigma^{a}\sigma^{b}\nonumber \\ & & +\, i \left( \frac{\lambda
 \mathtt{v}} {2\hbar} \right)^{3} \sum_{a,b,c} \int_{-\infty}^{\infty}
 dt \int_{-\infty}^{t} dt^{\prime} \int_{-\infty}^{t^{\prime}}
 dt^{\prime\prime} :\!J_{L}^{a} \left(t\right)\!:\; :\!J_{L}^{b}
 \left(t^{\prime} \right)\!:\; :\!J_{L}^{c} \left(t^{\prime\prime}
 \right)\!: \sigma^{a} \sigma^{b} \sigma^{c}.
\label{eq:Ukondo3order}
\end{eqnarray}
Due to the QEC evolution, only some of these terms are kept after the
syndrome is extracted [see Eq.~(\ref{eq:extended-evolution})].  For
clarity, let us assume that we know from the syndrome that a phase
flip has occurred. Hence, we must truncate the evolution operator to
reflect this fact and apply the recovery operation (in this case
multiply by $\sigma^{z}$), yielding
\begin{eqnarray} 
v_{z} & \approx & -\frac{i\lambda \mathtt{v}} {2\hbar}
\int_{-\infty}^{\infty} dt :\!J_{L}^{z} \left(t\right)\!: -
i\left(\frac{\lambda \mathtt{v}} {2\hbar} \right)^{2}
\int_{-\infty}^{\infty} dt \int_{-\infty}^{t} dt^{\prime} \left[
:\!J_{L}^{x} \left(t\right)\!:\; :\!J_{L}^{y}
\left(t^{\prime}\right)\!: - :\!J_{L}^{y} \left(t\right)\!:\;
:\!J_{L}^{x} \left(t^{\prime}\right)\!: \right] \nonumber \\ & & +\, i
\left(\frac{\lambda \mathtt{v}} {2\hbar} \right)^{3} \sum_{a}
\int_{-\infty}^{\infty} dt \int_{-\infty}^{t} dt^{\prime}
\int_{-\infty}^{t^{\prime}} dt^{\prime\prime} \left[
:\!J_{L}^{a}\left(t\right)\!:\; :\!J_{L}^{a}\left(t^{\prime}
\right)\!:\; :\!J_{L}^{z} \left(t^{\prime\prime}\right)\!: +
:\!J_{L}^{z}\left(t\right)\!:\;
:\!J_{L}^{a}\left(t^{\prime}\right)\!:\;
:\!J_{L}^{a}\left(t^{\prime\prime}\right)\!: \right. \nonumber \\ & &
-\, \left. :\!J_{L}^{a}\left(t\right)\!:\; :\!J_{L}^{z}
\left(t^{\prime}\right)\!:\;
:\!J_{L}^{a}\left(t^{\prime\prime}\right)\!: \right].
\label{eq:SzsigmazKondo}
\end{eqnarray}
\end{widetext}
Now, we integrate over a small frequency shell $\left[ \Lambda -
\delta\Lambda,\Lambda \right]$ and invoke the OPE. The result is a 
renormalization of the coupling $\lambda$ by an infinitesimal composed of
quadratic and cubic terms,
\begin{eqnarray}
\frac{d\lambda} {d\ell} & = & \lambda^{2} - \frac{k}{2}
\lambda^{3}.\end{eqnarray}
The resulting running coupling $\lambda(\ell)$ can be used to improve
the results of our bare perturbation theory. For that purpose, we
integrate the beta function from the bare cutoff until
$\Delta^{-1}$. For the case of a small number of channels, we obtain a
renormalized coupling of the form
\begin{equation} 
\lambda^\ast \approx \frac{\lambda} {1-\lambda \ln \left| \Lambda
\Delta \right|}.
\end{equation}
Although the RG flow goes toward the strong coupling limit, we do not
integrate the beta function all the way to zero frequency. Thus, if
the renormalized coupling $\lambda^\ast$ is still a small parameter,
it replaces $\lambda$ leading to the first-order \textit{renormalized}
evolution
\begin{equation}
v_{z} \approx - \frac{i\lambda^\ast \mathtt{v}} {2\hbar}
\int_{-\infty}^{\infty}dt :\!J_{L}^{z} \left(t\right)\!: \sigma^{z}.
\end{equation}
%

\subsubsection{Quantum frustrated system}

Correlations are not necessarily malignant to the computer's behavior.
This is illustrated by our second example: a quantum frustrated
environment \cite{NNB+03,NCN+05,NGN05}. Consider the case of three
independent Abelian ohmic baths coupled as in Eq.~(\ref{eq:Skondo}),
but with the OPE
\begin{equation}
:\!J_{L}^{a}\left(t\right)\!:\; :\!J_{L}^{b}
\left(t^{\prime}\right)\!: \;\to\; - \frac{\delta^{ab}}
{2\mathtt{v}^{2}\left( t -t^{\prime}\right)^{2}} \;.
\end{equation}
Following precisely the same methodology of the previous example, we
obtain the beta function
\begin{equation}
\frac{d\lambda}{d\ell} = -\frac{1}{2}\lambda^{3},
\end{equation}
which leads to the renormalized coupling
\begin{equation}
\lambda^\ast \approx \frac{\lambda} {\sqrt{1+2\lambda^{2} \ln \left|
\Lambda\Delta \right|}}.
\end{equation}
A quantum frustrated system has the remarkable property of asymptotic
freedom. Hence, even very large bare couplings flow towards a
perturbative regime. The physical reason behind this is the lack of a
pointer basis \cite{Zur81}, thus effectively decoupling the qubit from
its surroundings \cite{NCN+05}. This phenomena can also be understood
as self-inflicted $\pi$-pulse decoupling working at the cutoff
frequency $\Lambda$ \cite{VKL99,SV06}.

If the three coupling constants have different bare values, then the
flow stops at some finite frequency since two of the couplings will
flow to zero before the third. In other words, there will be a pointer
basis. In a quantum computer protected by QEC, however, we are
effectively stopping the flow at a finite frequency. Hence, the effect
described in the previous paragraph is relevant even for large
anisotropic couplings.

\subsection{Probability of a faulty path}
\label{sec:probfaultypath}

Now that we have obtained a reasonable approximation to the evolution
operator at each QEC step, we can turn to the problem of evaluating how
much protection QEC yields at long times. The simplest quantity to
calculate is the probability of finding a particular history of
syndromes, Eq.~(\ref{eq:hisprob}).
Using Eq.~(\ref{eq:v-evolution-reno})
and the known commutation relations of the
$f_{\alpha}$ operators, we in general can write that
\begin{equation}
\Upsilon_{{\mathbf{w}}}^{\dagger} \Upsilon_{{\mathbf{w}}} =
\upsilon_{w_{N}}^{2} \big( N\Delta, \left(N - 1\right)
\Delta\big)...\upsilon_{w_{1}}^{2} \big(\Delta,0\big),
\end{equation}
and define
\begin{equation}
\upsilon_{w}^{2} \big(\Delta,0\big) \approx \sum_{ij}
\frac{\lambda_{\alpha_{i}}^\ast \lambda_{\alpha_{j}}^\ast} {4 \hbar^{2}}
\int_{0}^{\Delta} dt_{1}\, dt_{2}\, f_{\alpha_{i}}^{\dagger}
\left({\bf x}_{i},t_{1}\right) f_{\alpha_{j}} \left({\bf x}_{j},t_{2}
\right).
\end{equation}
We now can evoke Wick's theorem once again to separate the intra- and
inter-hypercube contributions to the probability: The quantum average
$\mathcal{P} \left( \Upsilon_{{\mathbf{w}}} \right) \approx
\left\langle \varphi_{0} \right| \Upsilon_{{\mathbf{w}}}^{\dagger}
\Upsilon_{{\mathbf{w}}} \left| \varphi_{0} \right\rangle $ can be written
as a sum of all possible pair contractions. It is convenient to
separate the sum into two distinct parts. 

First, the sum of all pair contractions in the same hypercube gives
the stochastic error probability of a qubit, that we defined in
Eq.~(\ref{eq:localp}), namely,
\begin{eqnarray}
\label{eq:localerrorprop} 
\epsilon_{\alpha} & = & \left\langle \varphi_{0} \right|
\upsilon_{\alpha}^{2} \left({\bf x}_{1}, \lambda_{\alpha}^\ast \right)
\left| \varphi_{0} \right\rangle \nonumber \\ & = &
\left(\frac{\lambda_{\alpha}^\ast} {2 \hbar}\right)^{2} \int_{0}^{\Delta}
dt_{1} dt_{2} \langle f_{\alpha}^{\dagger}\left({\bf x},t_{1}
\right)f_{\alpha} \left({\bf x},t_{2}\right)\rangle,
\end{eqnarray}
where we used again that for $\left| {\bf x}-{\bf y} \right| > \xi$
and $\left| t_{1}-t_{2} \right| < \Delta$, we have $\left\langle
f_{\alpha}^{\dagger} ({\mathbf{x}},t_{1}) f_{\alpha}
({\mathbf{y}},t_{2}) \right\rangle \approx0$. Note that when we
calculated $\lambda^\ast$ we already summed intra-hypercube pair
contractions; however, these were contractions on the same Keldysh
branch [see Fig.~\ref{fig:Keldysh}(a)] and therefore are related to
the wave function amplitude. Equation (\ref{eq:localerrorprop})
corresponds to pair contractions between two distinct Keldysh branches
[see Fig.~\ref{fig:Keldysh}(b)], hence it gives the probability of
that evolution. With this two-step procedure, we sum up the most
relevant contributions to the probability within a hypercube.

Second, we sum contractions between hypercubes. For each possible
syndrome outcome we define the operators
\begin{equation}
F_{0}\left({\bf x},0\right) = 1 - \frac{\sum_{\alpha}
\left(\lambda_{\alpha}^\ast \Delta / 2 \hbar\right)^{2}} {1 -
\sum_{\alpha} \epsilon_{\alpha}}: \left| f_{\alpha} \left({\bf
x},0\right) \right|^{2}:
\end{equation}
and
\begin{equation}
F_{\alpha} \left({\bf x},0\right) = \frac{1} {\epsilon_{\alpha}}
\left( \frac{\lambda_{\alpha}^\ast\Delta} {2 \hbar}\right)^{2} :\left|
f_{\alpha}\left({\bf x},0\right) \right|^{2}:,
\end{equation}
where $:\, :$ stands for normal ordering with respect to the
environment ground state (see Appendix
\ref{sec:low-frequency-contribution}). We use these operators to
express the remaining pair contractions of each hypercube in the
probabilities, namely,
\begin{equation}
\upsilon_{0}^{2}\left( {\bf x},\Delta,0 \right) \approx \left( 1 -
\sum_{\alpha} \epsilon_{\alpha} \right) F_{0} \left( {\bf x},0 \right)
\label{eq:no-error-prob}
\end{equation}
and
\begin{equation}
\upsilon_{\alpha}^{2}\left({\bf x},\Delta,0\right) \approx
\epsilon_{\alpha} \left[ 1 + F_{\alpha} \left({\bf x},0 \right)
\right].
\label{eq:error-prob}
\end{equation}

Equations (\ref{eq:no-error-prob}) and (\ref{eq:error-prob}) are the
final ingredients needed to evaluate the probability of a particular
history of syndromes, Eq.~(\ref{eq:hisprob}). The remarkable aspect of
these equations is that they provide a very elegant reorganization of
the perturbation series. They were tailored to separate the local
contribution, $\epsilon_{\alpha}$, from the long-distance, long-time
components of the noise, $F_{\alpha}$. The high-frequency part gives
rise to the stochastic noise that is well discussed in the QEC
literature. We rewrote the rest of the series taking into account the
unusual non-unitary driven dynamics of QEC. The only remaining issue is
to evaluate the stability of the perturbation expansion in the {\it
renormalized} coupling $\lambda^\ast$.

In Sec.~\ref{sec:the-problem-of} we discussed how the scaling
dimension of an operator is important when studying a perturbative
expansion. The same argument holds when evaluating the protection
yielded by QEC in a correlated environment. If the scaling dimension
of $f_{\alpha}$ is $\delta_{\alpha}$, then $\dim F_{\alpha} = 2
\delta_{\alpha}$ (see Appendix \ref{sec:scaling-dimension-of}). Hence,
the original criterion for the validity of the perturbative expansion
in $\lambda$, $D + z - \delta_{\alpha} < 0$, becomes
\begin{equation}
D + z - 2 \delta_{\alpha} < 0
\label{eq:scalingequation}
\end{equation} 
once the expansion in $\lambda^\ast$ is adopted. Note the factor of
$2$ in this equation caused by QEC.

Whenever Eq.~(\ref{eq:scalingequation}) is satisfied, the long-range
correlations will produce small corrections to the stochastic error
probability. Below, we illustrate this point with an example.

\subsubsection*{Probability of a ``flawless'' evolution.}

Consider the case of a non-Markovian noise model with only one type of
error (phase flips, for instance). For simplicity, assume that no
spatial correlations exist ($D=0$). Hence, we can consider each qubit
separately and do not have to worry about the spatial structure of the
quantum computer. We also assume a two-point correlation function of
the form
\begin{equation}
\left\langle f \left( {\bf x},t_{1 }\right)\, f \left( {\bf y},t_{2}
\right) \right \rangle = \frac{1}{2} \left( \frac{\tau_{0}} { \left|
t_{1} - t_{2}\right|} \right)^{2\delta/z}\delta_{\bf{x},\bf{y}},
\end{equation}
where $\tau_{0}$ is a constant with the dimension of time. How do
these long-range correlations change the probability of a flawless
evolution of a qubit after $N \!\gg\! 1$ QEC steps? To answer this
question, we evaluate
\begin{eqnarray}
\mathcal{P} \left( \Upsilon_{{0}} \right) & \approx & \left\langle
 \varphi_{0} \right| \prod_{j=0}^{N-1} \upsilon_{0}^{2} \left( {\bf
 x}_{i},j\Delta \right) \left| \varphi_{0} \right\rangle \nonumber \\
 & \approx & (1-\epsilon)^{N} \left\langle \varphi_{0} \right|
 \prod_{j=0}^{N-1}F_{0} \left( {\bf x_{i}},j\Delta \right) \left|
 \varphi_{0} \right\rangle.
\label{eq:Pupslon0}
\end{eqnarray}
Assuming $\epsilon, \lambda^\ast \ll 1$, we can rewrite the
probability as
\begin{widetext}
\begin{eqnarray}
\mathcal{P} \left( \Upsilon_{{0}} \right) & \approx & 
e^{-N\epsilon} \left\langle \varphi_{0} \right| T_{t} \exp \left\{
-\frac{\left[ \lambda^\ast \Delta/ \left( 2 \hbar\right) \right]^{2}}
{1-\epsilon} \int_{0}^{N\Delta} \frac{dt} {\Delta} :\left|
f\left(t\right)\right|^{2}: \right\} \left| \varphi_{0} \right\rangle \nonumber
\\ & \approx & 
e^{-N\epsilon} \left\{ 1 + \frac{\left[ \lambda^\ast \Delta/ \left( 2
 \hbar \right)\right]^{4}} {(1-\epsilon)^{2}} \int_{0}^{N\Delta}
 \frac{dt_{1}} {\Delta} \int_{0}^{t_{1}} \frac{dt_{2}} {\Delta}
 \frac{\tau_{0}^{4\delta/z}} {\left( t_{1}-t_{2} \right)^{4\delta/z}}
 + \ldots \right\} \nonumber \\ & \approx & e^{-N\epsilon} \left\{ 1 +
 \frac{\left[\left( \lambda^\ast\Delta/ 2 \hbar \right)\right]^{4}}
 {(1 - \epsilon)^{2}} \frac{(\tau_{0}/\Delta)^{4\delta/z}
 N^{2(1-2\delta/z)}} {2(1 - 2\delta/z) (1-4\delta/z)} + \ldots
 \right\},
\label{eq:probupslon0g}
\end{eqnarray}
\end{widetext}
where we have kept only the leading term. There are two simple limits:

(i) If $z<2\delta$, the corrections become increasingly irrelevant as
$N$ grows. The stochastic probability in the limit of large $N$ is
given by $\mathcal{P} \left( \Upsilon_{{0}} \right) \approx
e^{-N\epsilon}$ and the correction due to correlations are small.

(ii) The tipping point is $z=2\delta$. By summing the subset of
dominant terms
\begin{equation}
\int_{0}^{N\Delta} \frac{dt_{1}}{\Delta}... \int_{0}^{t_{2j}}
\frac{dt_{2j+1}}{\Delta} \prod_{i=1}^{j} \left\langle : \left|
f\left(t_{2i-1} \right) \right|^{2}\!:\; :\!\left| f \left(t_{2i}
\right)\right|^{2}: \right\rangle,
\end{equation}
we obtain
\begin{equation}
\mathcal{P} \left(\Upsilon_{{0}}\right) \approx e^{-N\epsilon}
\frac{1}{1-\frac{\left(\lambda^\ast\Delta/\hbar\right)^{4}}
{\left(1-\epsilon\right)^{2}} \ln N}.
\end{equation}
This signals a problem with the perturbative expansion when $N \!\approx\!
\exp {\left( \hbar\frac{1-\epsilon}{\lambda^\ast\Delta}
\right)^{2}}$. For times larger than $\Delta \exp 
{\left(
\hbar\frac{1-\epsilon}{\lambda^\ast\Delta }\right)^{2}}$, correlations
substantially change the probability.

\subsection{Residual decoherence}
\label{sec:residdecohere}

In addition to the probability of a given syndrome history, we also
identified the residual decoherence, Eq.~(\ref{eq:rdmlqubit}), as a
fundamental quantity to QEC. The reason is that the noise models that
we consider do not satisfy the Lafflame-Knill condition for perfect
error correction \cite{KL97}, as is the case for most physically
relevant decoherence mechanisms. Hence, it may not be safe to ignore
these high-order events in the coupling $\lambda$.

It is straightforward to develop a calculation for the density matrix
along the same lines used for the syndrome history probability. After
separating the intra- and inter-hypercube contributions, the
perturbative expansion is reorganized using the renormalized coupling
$\lambda^\ast$. The result is exactly the same as for the case of the
probability: If $D+z-2\delta<0$, the perturbation theory in
$\lambda^\ast$ is stable and the analysis of the residual decoherence
done with the corresponding stochastic model is a good approximation
of the true quantum result. We revisit the example used in
Sec.~\ref{sec:faultpath} to make this point clear.

\subsubsection*{Decoherence of a ``flawless'' evolution.}

For this example, we assume an environment that can only introduce
phase flip errors in the computer. As we discussed in
Sec.~\ref{sec:intro}, for this error model we can use the simple
3-qubit code. However, unlike the calculation of the probability of a
flawless evolution, we now make some assumptions about the spatial
structure of the computer: We consider for simplicity that each
logical qubit is composed of three \emph{adjacent} physical
qubits. The encoding and decoding are described in
Fig.~\ref{fig:3-qubit-code}.

Following Ref.\ \onlinecite{NB06}, we write the evolution operator for
a particular logical qubit in a QEC cycle as
\begin{eqnarray}
\mathtt{w}_{0}\left(0,\bar{{\bf x}}_{0}\right)&=&\upsilon_{0}\left({\bf
x}_{1},0\right)\upsilon_{0}\left({\bf
x}_{2},0\right)\upsilon_{0}\left({\bf
x}_{3},0\right).
\label{evol-log-1}
\end{eqnarray}
By expanding Eq.~(\ref{evol-log-1}) in powers of $\lambda$, we obtain
\begin{widetext}
\begin{eqnarray}
\mathtt{w}_{0}\left(0,\bar{{\bf x}}_{0}\right) & = & 1 - \left(
\frac{\lambda}{2\hbar}\right)^{2} \sum_{j} \int_{0}^{\Delta} dt_{1}
\int_{0}^{t_{1}} dt_{2}f \left({\bf x}_{j},t_{1}\right)f\left({\bf
x}_{j}, t_{2} \right) \nonumber \\ & + & i \left(
\frac{\lambda}{2\hbar} \right)^{3} \int_{0}^{\Delta} dt_{1}
\int_{0}^{\Delta} dt_{2} \int_{0}^{\Delta} dt_{3} f \left({\bf
x}_{1},t_{1} \right) f \left({\bf x}_{2},t_{1} \right) f \left({\bf
x}_{3},t_{1} \right) \bar{Z},
\end{eqnarray}
%
%
where $\bar{Z}$ is the logical phase flip for that particular logical
qubit. Note that the third order term keeps the logical qubit inside
the logical Hilbert space \cite{NB06} and therefore is not corrected
by the QEC code.

We choose to evaluate the most off-diagonal term of the reduced
density matrix,
%
%
\begin{eqnarray}
\rho_{\vec{\uparrow},\vec{\downarrow}}\left(\Upsilon_{{\mathbf{0}}}\right)
& = &\frac{\left\langle \varphi_{0}\right|\left[\left\langle
\psi_{0}\right| \prod_{j=N-1}^{0} \prod_{k=1}^{M}
\mathtt{w}_{0}^{\dagger}\left(j\Delta,\bar{{\bf x}}_{k} \right) \left|
\vec{\downarrow} \right\rangle \left\langle \vec{\uparrow} \right|
\prod_{j=0}^{N-1} \prod_{k=1}^{M} \mathtt{w}_{0}
\left(j\Delta,\bar{{\bf x}}_{k}\right)\left|\psi_{0}\right\rangle
\right]\left|\varphi_{0}\right\rangle }{\left\langle
\varphi_{0}\right|\left\langle \psi_{0}\right| \prod_{j=0}^{N-1}
\prod_{k=1}^{M} \mathtt{w}_{0}^{2} \left({\bar{\bf
x}}_{k},j\Delta\right)\left|\psi_{0}\right\rangle
\left|\varphi_{0}\right\rangle },
\end{eqnarray}
%
%
where $\vec{\uparrow} = \left| \uparrow ... \uparrow \right\rangle$
and $\vec{\downarrow} = \left|\downarrow ... \downarrow \right\rangle$
denote the state of the physical qubits, $\bar{\bf{x}}_{k}$ is
labeling $M$ logical qubits, and $N$ is the total number of QEC steps.

After integrating all the modes inside a hypercube, we define a
renormalized coupling $\lambda^\ast$ and a local error probability
$\epsilon$. Finally, we evoke again Wick's theorem to write
%
%
\begin{equation}
\rho_{\vec{\uparrow},\vec{\downarrow}}
\left(\Upsilon_{0}\right)=\left\langle \psi_{0}|\vec{\downarrow}
\right\rangle \left\langle \vec{\uparrow}|\psi_{0}\right\rangle
\frac{1-A-NM\epsilon^{3}-\epsilon^{4} \left(
\frac{\lambda^\ast\Delta}{2\hbar} \right)^{4}\sum_{\bar{{\bf x}},
\bar{{\bf y}}} \int_{0}^{N\Delta} dt_{1} \int_{0}^{t_{1}}dt_{2}
\left\langle :\!f^{2} \left(\bar{{\bf x}},t_{1}\right)\!:\; :\!f^{2}
\left(\bar{{\bf y}},t_{2}\right)\!: \right\rangle +...}
{1-A+NM\epsilon^{3} + \epsilon^{4} \left(
\frac{\lambda^\ast\Delta}{2\hbar} \right)^{4} \sum_{\bar{{\bf x}},
\bar{{\bf y}}} \int_{0}^{N\Delta} dt_{1} \int_{0}^{t_{1}} dt_{2}
\left\langle :\!f^{2} \left(\bar{{\bf x}},t_{1} \right)\!:\; :\!f^{2} \left(
\bar{{\bf y}},t_{2} \right)\!: \right\rangle +...},
\end{equation}
where $A$ is a number proportional to $\epsilon$ and $\lambda^\ast$.
Hence, for $\epsilon,\lambda^\ast\ll1$, this simplifies to
\cite{EfficacyQEC}
\begin{equation}
\rho_{\vec{\uparrow},\vec{\downarrow}} \left(\Upsilon_{0} \right)
\approx \left\langle \psi_{0} | \vec{\downarrow} \right\rangle
\left\langle \vec{\uparrow}|\psi_{0}\right\rangle \left[ 1 - 2NM
\epsilon^{3}-2\epsilon^{4} \left( \frac{\lambda^\ast\Delta}{2\hbar}
\right)^{4} \int \!d{\bf x} \int \!d{\bf y} \int_{0}^{N\Delta}
\!dt_{1} \int_{0}^{t_{1}} \!dt_{2} \left\langle :\!f^{2} \left(
\bar{{\bf x}},t_{1} \right)\!:\; :\!f^{2} \left(\bar{{\bf y}},t_{2}
\right)\!: \right\rangle + \ldots \right].
\label{eq:rhoflawless}
\end{equation}
\end{widetext}
If we now recall the two-point correlation function of
Eq.~(\ref{general2pointfunction}), it becomes clear that the
corrections due to correlations are relevant when $D+z>2\delta$.

\subsection{Relation to the work of Aharonov, Kitaev, and Preskill}

The study of correlated noise has been a central problem for quite
some time. Among the most recent advances is a paper by Aharonov,
Kitaev, and Preskill (AKP) \cite{AKP06}. Using a method completely
different from ours, AKP proved that: For a computer where qubits are
interacting through an instantaneous interaction of the form
$\lambda^{2} / \Delta x^{2\delta}$, it is possible to prove resilience
for $\lambda < \lambda_{c}$ and $D-2\delta < 0$. The key distinction
between the work of AKP and ours is the instantaneous nature of their
interaction. Hence, while in our work each qubit is inside a distinct
hypercube, for AKP they are all contained in a single hypercube. There
is however a trade-off. Since their interaction is instantaneous and
perfect error correction is assumed, there is no propagation of errors
in time through the gauge field of the environment. Hence,
effectively, AKP are considering a model with $z=0$. As a result, our
Eq.~(\ref{eq:scalingequation}) holds in the case they analyzed as
well.

\section{Threshold Theorem as a Quantum Phase Transition}
\label{sec:QPT}

%
\begin{figure*}
\includegraphics[width=12cm]{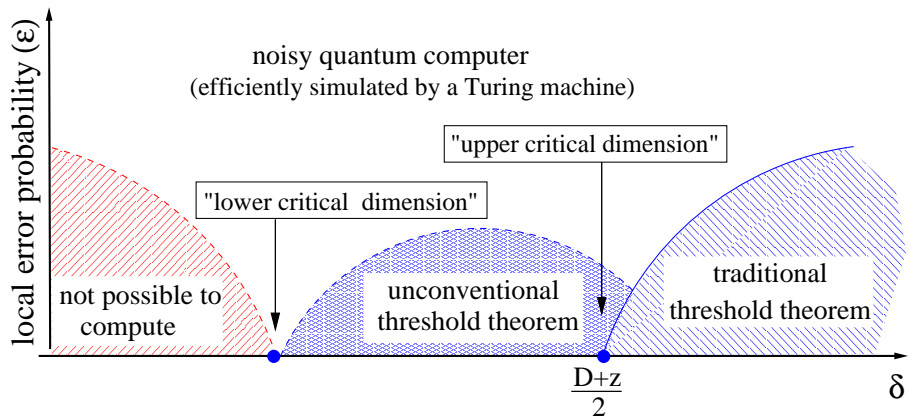}
\caption{(Color online) Phase diagram of a quantum computer running
QEC.  The parameter $\delta$ is the scaling dimension of the
environment operator, D is the dimensionality of the computer, and z
is the dynamical exponent of the environment [see discussion preceding
Eq.~(\ref{eq:scalingequation})]. In the red phase, qubits and
environment are strongly entangled causing strong decoherence. In the
light blue phase, QEC keeps the qubits and environment disentangled,
making computation possible.  }
\label{fig:phase-diagram}
\end{figure*}
%

The main result of fault-tolerant quantum computation is the threshold
theorem. The theorem states that if a stochastic error probability
$\epsilon$ is smaller than a critical value $\epsilon_{c}$, then the
introduction of an additional layer of concatenation improves the
protection of the information. Hence, for a fixed $\epsilon$, it is
possible to sustain a quantum computation for any desire time at the
cost of some reasonable additional hardware overhead.

Even though quantum computation and QEC are out-of-equilibrium
problems, it is intuitive to talk about different phases in the
computer-environment parameter space. Along this line of thought, each
phase corresponds to a distinct steady state. A natural choice for an
order parameter is that given by the entanglement among the qubits and
the environment. We summarize our thinking in Fig.~\ref{fig:phase-diagram}, 
where we present a schematic phase diagram for a quantum computer running
QEC.

For stochastic noise models, such an idea was explored by Aharonov
\cite{Aha00}. Following that work, we can separate the behavior of the
computer into two distinct regimes:

(i) For $\epsilon\!<\!\epsilon_{c}$, the computer components can
maintain large entanglement through fault-tolerant procedures, which
in turns means that the computer and the environment are weakly
entangled.  Hence, due to this large internal entanglement, the
quantum computer departs from the classical computer model.
We can formalize these
remarks by remembering that QEC tries to keep the system in the ``steady
state" described by the reduced density matrix. In order to keep the
notation simple, let´s take the ideal computer state as a pure state,
\begin{equation}
\rho\left(t\right) = \left| \psi\right\rangle \left\langle
\psi\right|,
\end{equation}
with $\left| \psi \right\rangle \!=\! \sum_{i} \alpha_{i}
\left(t\right) \left| i\right\rangle$ expressed in terms of the
computational basis $\{\left|i\right\rangle\}$. As consequence, it has
a reduced entropy $S \approx 0$. In this case, if we look at the full
Hilbert space (that is, before tracing out the environment), we find
the tensor state
\begin{equation}
\left| \Psi \right\rangle \approx \left| \psi \right\rangle \otimes
\left| \varphi_{\mbox{environment}} \right\rangle \;.
\label{eq:freefixedpoint}
\end{equation}

(ii) For $\epsilon\!>\!\epsilon_{c}$, the computer components are
weakly entangled and, therefore, can be efficiently simulated by a
Turing machine. In other words, the computer density matrix no longer
represents a pure state, but rather a statistical mixture. Thus, the
computer components are strongly entangled with the environment. This
corresponds to a steady state with a large reduced entropy (in the
limit of $\epsilon\to1$, $S\approx N\ln2$, with $N$ the number of
qubits).

In such a description, we see that $\epsilon$ plays a role analogous
to an effective temperature \cite{spinlatt}.  Hence, the threshold
theorem defines a phase transition from a high-temperature phase,
where qubits are independent from each other, to a low-temperature
phase, where quantum coherence and entanglement are possible
\cite{Aha00}. This also sheds new light on the role of periodic
measurements in QEC: They can be seen as a refrigeration that extracts
entropy from the computer (very much like the Schulman-Vazirani
initialization procedure \cite{SV99} or the transfer of entanglement
to fresh ancillas \cite{Preskill-notes}). If the entropy production in
the computer is below a certain level, then the computer can be kept
in its ``low-temperature'' phase.

Our analysis of correlated noise also fits perfectly into this
description. The dimension criterion provided by
Eq.~(\ref{eq:scalingequation}) is the hallmark of a quantum phase
transition \cite{Sac99}. For $D+z<2\delta$, $V$ can only produce small
corrections to the stochastic error model. The steady state of the
system is therefore given by Eq.~(\ref{eq:freefixedpoint}). There is a
clear separation of scales and the threshold theorem holds as it
is. Conversely, for $D + z > 2\delta$, there is no clear separation of
scales. The computer and the environment become increasingly entangled
and the system is driven towards a different steady state. Such a
state is probably distinct from the ``high temperature'' one and it is
likely that it is characterized by a smaller residual entropy.

This does not mean that for $D + z > 2\delta$ it would not be possible
to perform quantum computation. It only means that the threshold
theorem as we stated it does not hold. It is conceivable that some
different derivation of the theorem exists in this case. In this
sense, $D + z = 2\delta$ defines what is usually referred to as the
upper critical dimension of the model (see Appendix
\ref{sec:Perturbative-expansion-in}). Below the upper critical
dimension, there can be substantial corrections to the steady state
given by Eq.~(\ref{eq:freefixedpoint}), but it may still be possible
to prove resilience. The question that remains open is whether a lower
critical dimension exists, namely, a criterion for $V$ that would tell
us when it is impossible to perform long-time quantum computation.


\section{Summary and Conclusions}
\label{sec:conclusions}

Most previous discussions of QEC have used the quantum master equation
and quantum dynamical semi-groups \cite{BP02}. This is a very natural
approach: The computer is the object of interest; hence, one starts
the discussion by integrating out the environmental degrees of
freedom.  However, the price paid in this approach is that some
simplification is needed in order to derive the quantum master
equation \cite{BP02,ALZ06}. The usual assumption is the Born-Markov
approximation \cite{ALZ06}. In that case, it is natural to define an
error probability for a given qubit, and a discussion in terms of
error models naturally follows \cite{NC00,AGP06}. The situation is much less
clear when the Born-Markov approximation cannot be justified
\cite{TB05,DL05}. In this case, temporal and spatial correlations can
build up and completely destroy the notion of the probability of an
error.

A key characteristic of the discussion here is that we do not try to
use a quantum master equation. Rather, we follow the approach put
forward by Schwinger and Keldysh \cite{Sch61,Kel65,mahan} to study out
of equilibrium systems. The main conceptual difference is that we
trace the environmental degrees of freedom only at the very last step
of the calculation. Hence, we can make the most of the unitary
evolution of a quantum mechanical system.

Following this ``Schwinger-Keldysh'' approach, we discussed the
evolution of a quantum computer operated with fast and slow gates. On
the one hand, for fast gates the microscopic Hamiltonian is the one
relevant for the evolution of the computer,
Eq.~(\ref{eq:fastgates}). On the other hand, for slow gates we
demonstrated that a suitable effective Hamiltonian,
Eq.~(\ref{eq:bigF}), can be used to provide an upper bound for the
discussion of decoherence. With this effective Hamiltonian, the
notation can be unified, and both cases treated simultaneously.  We
derived two formal expressions that quantify the evolution of the
computer under QEC in a correlated environment: (i) the probability of
a given syndrome history, Eq.~(\ref{eq:hisprob}), and (ii) the reduced
density matrix of the computer, Eq.~(\ref{eq:rdmlqubit}).

In order to fully use standard QEC theory, we introduced the important
assumption of ``hypercubes'', that is a minimum spatial distance
between qubits, Eq.~(\ref{xi-def}), in order to allow the definition
of an error probability for a single qubit. With this ``hypercube
assumption'', it is straightforward to use Wick's theorem to separate
the environmental modes into intra- and inter-hypercube parts. The
intra-hypercube component defines the error probability, while the
inter-hypercube part is tracked by an operator acting on the
coarse-grained scale of the hypercubes. As examples, we treated a
generalization of the spin-boson model and a quantum frustrated model.

All the pieces are put together when we explicitly calculate the
probability of a syndrome history (Sec.~\ref{sec:probfaultypath}) and
associated residual decoherence (Sec.~\ref{sec:residdecohere}). The
main result is cast as a dimensional criterion,
Eq.~(\ref{eq:scalingequation}). Finally, we discuss the parallels
between the threshold theorem and a quantum phase transition. A
qualitative description of the possible fates of a quantum computer as
a function of noise strength and degree of correlation is given in
Fig.~\ref{fig:phase-diagram}.

There are several clear directions in which our results could be
extended or improved. First, it would obviously be desirable to relax
the hypercube assumption introduced in Sec.\
\ref{sec:hypercube}. There is nothing intrinsic to our approach which
makes this assumption necessary. Yet, progress without it seems much
more difficult: The notion of a local error probability during a
single QEC cycle becomes problematic, making the connection with
analysis based on error models, such as the usual derivation of the
threshold theorem, unclear.

Second, non-instantaneous gate operation is clearly a delicate
issue. By using a bound (Sec.\ \ref{sec:upperbounds}), we are able to
treat this case in the same way as the fast-gate case. Thus we derive
an upper bound for the local error probability together with the
dimensional criterion. If a more accurate value for the error
probability is desired, a specific error correction code as well as
the gates under consideration must be included in the
analysis. However, the scaling argument and resulting dimensional
criterion do not, in general, change.

Note that it is possible to change the dimensional criterion for the
better (but \textit{not} for the worse) by using the separation of
scales introduced by QEC. Particular pulse sequences can reduce
correlation at long times at the cost of increasing the local error
probability. One example was given in our previous work
\cite{NB06,NMB07}.

Finally, there may be a regime of parameters where, as indicated in
Fig.~\ref{fig:phase-diagram}, fault-tolerant quantum computation is
possible even though the presently known derivations of the threshold
theorem do not apply. By analogy with phase transition phenomenology,
there may be a lower critical dimension such that a more sophisticated
analysis than the one we present here shows that fault-tolerant
computation is possible for $\delta \!<\! (D+z)/2$. It would be very
interesting to show in any example that such is, or is not, the case.

Quantum Error Correction is one of the most interesting frameworks
which allows long quantum computations \cite{topqcomp}. Even though
QEC is widely accepted, it has been argued that it relies on a set of
unphysical assumptions \cite{AHH+02,AHH+04,ALZ06,Ali07}, namely: (i)
``fast'' measurements, (ii) ``fast'' gates, and (iii) describing
decoherence by error models. Although these are legitimate concerns,
it is now clear that they are not fundamental: First, in
Ref.~\onlinecite{DA07} DiVincenzo and Aliferis demonstrated that
resilient circuits can be constructed with slow measurements.  Second,
in the current paper, we have demonstrated that the fast gate
assumption is not critical for fault tolerance.  Finally, we have laid
the groundwork here for a theoretical framework that connects
microscopic Hamiltonians with error models in correlated
environments. From our results for the threshold theorem in
conjunction with those of AKP \cite{AKP06}, it is clear that a large
class of correlated environments are already properly treated within
the QEC framework.

\section*{Acknowledgments}

We thank C. Kane, D. Khveshchenko, and R. Plesser for useful
discussions. This work was supported in part by NSF Grants No.\ CCF
0523509 and No.\ CCF 0523603. E.R.M. acknowledges partial support from
the Interdisciplinary Information Science and Technology Laboratory
(I$^{2}$Lab) at UCF.

\appendix

\section{Absolute convergence of Dyson's~series}
\label{sec:absolute-convergence}

Dyson's series is absolutely convergent for any bound operator
evolving for any finite time \cite{OR00}. This is particularly simple
to see using the $\sup$ operator norm \cite{TB05},
\begin{equation}
\left| \left| A \right| \right| = \sup_{\Psi} \sqrt{\left\langle \Psi
\right| A^{\dagger} A \left| \Psi \right\rangle},
\end{equation}
where $\left| \left| \Psi \right| \right| = 1$. If $P = \int_{0}^{t}
dt^{\prime} \left| \left| V \left(t^{\prime}\right) \right| \right| <
\infty$, then the norm of the $m^{\rm th}$-order term in Dyson's
series is bounded by $P^{m}/m!$. Thus, using the convergence of the
exponential series, we find that Dyson's series is absolutely
convergent.

\section{\label{sec:Perturbative-expansion-in}Perturbative expansion 
in $\phi^{4}$~theory}

A classic example of a quantum phase transition is given by the
$\phi^{4}$ theory at criticality \cite{NO98}. The model is compactly
described by the Euclidean action
\begin{equation}
S =  \int_{0}^{L} d^{D}r \int_{0}^{\beta} d\tau \left[ \left(
\bigtriangledown_{r} \phi \right)^{2} + \left( \partial_{\tau}
\phi\right)^{2} + \lambda\phi^{4} \right] \;.
\end{equation}
The scaling dimension of the free field is usually defined as $\dim
\left[ \phi \right] = \nu/2$. If we expand the partition function in
powers of $\lambda$, it is simple to see that each order in the
perturbative expansion will have the power
$\lambda(L\beta)^{D+1-\nu}$. Hence, $D+1-\nu < 0$ is the criterion for
the irrelevance of the perturbation. The simplest way to see that is
to do power counting by rescaling space and time,
%
\begin{equation}
r \to br, \quad \tau \to b\tau, \quad \phi \to b^{-\frac{\nu}{2}}
\phi,
\end{equation}
which immediately gives
\begin{eqnarray}
S & = & b^{D-1-\nu} \int d^{D} r \int d \tau \left[ \left(
 \bigtriangledown_{r} \phi\right)^{2} +
 \left(\partial_{\tau}\phi\right)^{2} \right] \nonumber \\ & & +\,
 \lambda\, b^{D+1-2\nu} \int d^{D}r \int d\tau\, \phi^{4}.
\end{eqnarray}
One finds the scaling $\lambda \to \lambda b^{D+1-\nu}$, which is
valid at each order of the perturbative expansion. The criterion for
the irrelevance of the perturbation is $D+1-\nu<0$

There is one more important definition that this example provides.
Since the Gaussian action must be scale invariant, we automatically
see that for this example $\nu=D-1$. Hence, the criteria for the
irrelevance of $\lambda\phi^{4}$ term as a perturbation can be
rewritten as $3-D<0$. This defines the upper critical dimension for
the model as $d_{c}^{\rm upper}=4$ (three spatial and one
temporal). When a system is above its upper critical dimension, the
physics is controlled by the Gaussian action. However, when the system
is below its upper critical dimension, there are substantial
corrections to physical quantities when compared with the Gaussian
solution.

\section{Hilbert space of qubits}
\label{sec:hilbert}

Due to the state vector normalization, the Hilbert space of a qubit is
isomorphic to a three-dimensional sphere $S^{3}$: For a general state
$\left| \psi \right\rangle = \alpha \left| 1 \right\rangle + \beta
\left| 0 \right\rangle$, we have the constraint
\begin{equation}
\left( \mbox{Re}\,\alpha\right)^{2} + \left(
\mbox{Im}\,\alpha\right)^{2} + \left( \mbox{Re}\,\beta\right)^{2} +
\left(\mbox{Im}\,\beta\right)^{2} = 1.
\end{equation}
However, an overall phase is physically irrelevant and the correct
mapping is to the complex projective plane of complex dimension $1$,
\begin{equation}
S^{3}/U\left(1\right) \to \mathbb{CP}^{1}.
\end{equation} 
For the same reason, the Hilbert space of $n$ qubits is isomorphic to
$\mathbb{CP}^{2n-1}$. For the discussion of entanglement, there is a
particularly important subspace of this space. It is composed by the
direct product of each qubit Hilbert space minus an over all phase,
\begin{equation}
\left.\prod_{j=1}^{n} \mathbb{CP}_{\left(j\right)}^{1}
\right|_{\mbox{modulus phase}} \subset \mathbb{CP}^{2n-1},
\end{equation} 
where $j$ labels the $j^{th}$ qubit's Hilbert space. The dimension of
the subspace grows as $n-1$ while the dimension of the entire Hilbert
space grows as $2n-1$. Entangled states are defined as the
complementary set of this special subspace.

\section{Decoherence in the spin-boson model with ohmic dissipation}
\label{sec:SB-decoherence}

An example of a qubit coupled to an environment is the spin-boson
model with ohmic dissipation \cite{2stateRMP87,GPW99}, which was
intensively studied in the context of quantum computation
\cite{UNR95,RQJ02} even before quantum error correction was
introduced. In this model, a qubit evolves according to the
Hamiltonian
\begin{equation}
H = \int dx \left[ \left(\partial_{x}\phi\right)^{2} + \Pi^{2} \right]
+ \lambda\, \partial_{x} \phi\left(0\right) \sigma^{z},
\end{equation}
where $\phi$ is a chiral bosonic field, $\vec{\sigma}$ are Pauli
matrices that describe the qubit located at $x=0$, and $\lambda$ is
the environment-qubit coupling constant. If a qubit is prepared in an
initial state
\begin{equation}
\left|\psi\right\rangle = \alpha\left|\uparrow\right\rangle
+\beta\left|\downarrow\right\rangle,
\end{equation}
at large enough times, $\Lambda^{-1} \ll t \ll (k_BT)^{-1}$, its
density matrix evolves as
\begin{equation} 
\rho\left(t\right) = \left[\begin{array}{cc} \left|\alpha\right|^{2} &
\alpha\beta^\ast \, e^{-\lambda^{2}\ln\left(1+\Lambda t\right)}\\
\alpha^\ast \beta\, e^{-\lambda^{2}\ln\left(1+\Lambda t\right)} &
\left|\beta\right|^{2}
\end{array}\right],
\end{equation} 
with $\Lambda$ denoting the environment ultraviolet cutoff
frequency. Since states with either $\alpha$ or $\beta$ equal to zero
do not experience decoherence, they are called classical states. They
define a pointer basis. Conversely, any superposition state with
$\alpha,\beta \neq 0$ suffers decoherence and over a long time becomes
a statistical mixture of the classical states.

As one includes more qubits, the entries in the reduced density matrix
will decay faster as one moves away from the diagonal. In the case
where qubits are coupled to independent baths, it is simple to see
that the off-diagonal matrix elements decay as
\begin{equation}
\rho_{\vec{p},\vec{q}} \left( t\gg\Lambda^{-1} \right) = \rho_{0}\,
e^{-\lambda^{2} (p-q) \ln (1+\Lambda t)},
\end{equation}
where $p$ and $q$ are the total magnetization of the states $\vec{p}$
and $\vec{q}$, respectively \cite{UNR95}. The case of a common bath is
also straightforward, \cite{UNR95}, and the result for qubits
separated by a distance smaller than $\Lambda^{-1}$ is
\begin{equation}
\rho_{\vec{p},\vec{q}} \left( t\gg\Lambda^{-1} \right) \approx
\rho_{0}\, e^{-\lambda^{2} (p-q)^{2} \ln (1+\Lambda t)}.
\end{equation}

Some entangled states do not suffer decoherence (a singlet state, for
example). However, these correspond to a very special and small
decoherence-free subspace. In general, entangled states are made of
quantum superpositions and therefore have components in the
off-diagonal entries of the density matrix. Hence, studying
decoherence (the decay of the off-diagonal elements of the density
matrix) is essentially equivalent to studying how entanglement between
qubits is destroyed by interaction with the environment.

\section{Interaction picture}
\label{sec:Interaction-picture}

Since $\left[H_{0},H_{\rm QC}\right] = 0$, we can define the
interaction picture
\begin{eqnarray}
O(t) & = & e^{\frac{i}{\hbar}H_{0} t} R^{\dagger}(t)\, O\, R(t)\,
e^{-\frac{i}{\hbar}H_{0}t},\\ \left| \Psi (t) \right\rangle & = &
e^{\frac{i}{\hbar}H_{0}t} R^{\dagger}(t)\, \tilde{U}(t)\, \left|
\Psi(0) \right\rangle,
\end{eqnarray} 
where $\tilde{U}(t)$ is the exact evolution operator, defined as
\begin{equation} 
U(t) = T_{t}\, e^{-\frac{i} {\hbar} \int_{0}^{t} dt^{\prime}
H(t^{\prime})},
\end{equation}
and $|\Psi \rangle$ is the total state vector (computer plus
environment). Now, let us consider the time evolution of
$|\Psi\rangle$,
\begin{eqnarray}
\frac{d}{dt} \left| \Psi (t) \right\rangle & = & \frac{d}{dt}\,
 e^{\frac{i}{\hbar}H_{0}t}\, R^{\dagger}(t)\, \tilde{U}(t)\, \left|
 \Psi(0) \right\rangle \nonumber \\ & = & -\frac{i}{\hbar} V (t)
 \left| \Psi (t) \right\rangle.
\end{eqnarray}
Thus, we obtain the usual definition for the evolution operator in the
interaction picture
\begin{equation}
\tilde{U}(t) = e^{\frac{i}{\hbar}H_{0}t}\, R^{\dagger}(t)\,
\tilde{U}(t) = T_{t}\, e^{-\frac{i}{\hbar}\int_{0}^{t}dt^{\prime}\, V
(t^{\prime})}.
\end{equation}
%

\section{Low frequency contribution to the error probability}
\label{sec:low-frequency-contribution}

The simplest way to understand $F_{\alpha}$ is to write $f$ in its
frequency representation
\begin{eqnarray}
\lefteqn{ \upsilon_{\alpha} \left( {\bf x}_{1},\lambda_{\alpha}^\ast \right) 
 \approx  \lambda_{\alpha}^\ast \int_{0}^{\Delta} dt\, f_{\alpha}
 \left( {\bf x_{1}},t \right)} & & \nonumber \\ 
& \approx &
 \lambda_{\alpha}^\ast \int_{0}^{\Delta} dt \int_{0}^{\Lambda}
 d\omega\, e^{i\omega t} f_{\alpha} \left( {\bf x_{1}},\omega \right)
 \nonumber \\ 
& \approx & \lambda_{\alpha}^\ast \int_{0}^{\Delta} dt
 \left( \int_{0}^{\Delta^{-1}} d\omega + \int_{\Delta^{-1}}^{\Lambda}
 d\omega\ \right) e^{i\omega t} 
 f_{\alpha} \left( {\bf x_{1}},\omega \right) \nonumber \\ 
& \approx &
 \lambda_{\alpha}^\ast \int_{0}^{\Delta} dt \left[ f_{\alpha}^{>}
 \left({\bf x_{1}},t \right) + f_{\alpha}^{<} \left( {\bf x_{1}},0
 \right) \right],
\end{eqnarray}
where $<$ stands for frequencies smaller than $\Delta^{-1}$ and $>$
for the frequencies between $\Delta^{-1}$ and $\Lambda$. Thus, using
that $\left\langle f_{\alpha}^{<} f_{\alpha}^{>} \right\rangle = 0$,
we obtain
\begin{eqnarray}
\upsilon_{\alpha}^{2} \left({\bf x}_{1},\lambda_{\alpha}^\ast \right)
 & \approx & \left( \lambda_{\alpha}^\ast \right)^{2} \int_{0}^{\Delta}
 dt_{1}\, dt_{2}\, f_{\alpha}^{>\dagger} \left( {\bf x_{1}},t_{1}
 \right) f_{\alpha}^{>} \left({\bf x_{1}},t_{2} \right) \nonumber \\ &
 & +\, \left(\lambda_{\alpha}^\ast \Delta\right)^{2}
 f_{\alpha}^{<\dagger} \left({\bf x_{1}},0\right) f_{\alpha}^{<}
 \left( {\bf x_{1}},0 \right).
\end{eqnarray}
%

\section{Scaling dimension of $F_{\alpha}$}
\label{sec:scaling-dimension-of}

If the two-point correlation function of $f_{\alpha}$ can be expressed
as
\begin{equation}
\left\langle f_{\alpha} \left({\bf x}_{1},t_{1} \right)\, f_{\alpha}
\left({\bf x}_{2},t_{2} \right) \right\rangle \sim \mathcal{F} \left(
\frac{1}{(\Delta x)^{2\delta}},\frac{1}{(\Delta
t)^{2\delta/z}} \right),
\end{equation}
the scaling dimension of $f_{\alpha}$ is defined as $\dim
f_{\alpha}=\delta$. Using Wick's theorem,
\begin{widetext}
\begin{eqnarray}
\left\langle :\left|f_{\alpha} \left({\bf x}_{1},t_{1}\right)
 \right|^{2}\!:\; :\!\left| f_{\alpha} \left({\bf x}_{2},t_{2} \right)
 \right|^{2}: \right\rangle & = & \left\langle
 f_{\alpha}^{\dagger}\left({\bf
 x}_{1},t_{1}\right)f_{\alpha}^{\dagger}\left({\bf
 x}_{2},t_{2}\right)\right\rangle \left\langle f_{\alpha}\left({\bf
 x}_{1},t_{1}\right)f_{\alpha}\left({\bf
 x}_{2},t_{2}\right)\right\rangle \nonumber \\ & & +\, \left\langle
 f_{\alpha}^{\dagger}\left({\bf
 x}_{1},t_{1}\right)f_{\alpha}\left({\bf
 x}_{2},t_{2}\right)\right\rangle \left\langle f_{\alpha}\left({\bf
 x}_{1},t_{1}\right)f_{\alpha}^{\dagger}\left({\bf
 x}_{2},t_{2}\right)\right\rangle \nonumber \\ & = &
 2\left[\mathcal{F}\left(\frac{1}{\left(\Delta
 x\right)^{2\delta}},\frac{1}{\left(\Delta
 t\right)^{2\delta/z}}\right)\right]^{2}.
\end{eqnarray}

Therefore, $\dim F_{\alpha} = 2\delta$.
\end{widetext}
%



\begin{thebibliography}{68}
\bibitem[{sta()}]{static-imperfections}
\bibinfo{note}{Static imperfections can be dealt using randomization techniques
  \cite{KAS05}.}
\bibitem[{Eff()}]{EfficacyQEC}
\bibinfo{note}{The fact that the corrections in Eq.~(\ref{eq:rhoflawless})
  start at order $\epsilon^3$ is a direct demonstration of the efficacy of
  QEC.}
\bibitem[{spi()}]{spinlatt}
\bibinfo{note}{A spin lattice model is quite appropriate to illustrate the
  analogy. Suppose that we are looking at the 3-D Heisenberg spin Hamiltonian,
  $H = \frac{1}{2} \sum_{\left\langle ij \right\rangle} \vec{S}_{i} \cdot
  \vec{S}_{j}$, where $\left\langle ij \right\rangle$ denotes nearest neighbors
  spins. The ground state of this Hamiltonian is a singlet (anti-ferromagnetic
  phase), hence it exhibits a very strong entanglement among the spins. At low
  temperatures, this feature dominates the spin dynamics and the system is in
  its quantum phase. Conversely, at very high temperatures, all spin
  configurations are equally probable. The Hamiltonian cannot drive any
  coherent motion and the spins  effectively decouple (i.e, the system is in
  a paramagnetic phase).}
\bibitem[{top()}]{topqcomp} \bibinfo{note}{The other one is
topological quantum computation \cite{Kit97,KSV00}.}
\end{thebibliography}
\end{document}